\documentclass[a4paper,11pt]{article}
\pdfoutput=1

\usepackage{graphicx}
\usepackage{caption}
\usepackage{subcaption}
\usepackage[utf8]{inputenc}

\usepackage{fullpage}
\usepackage[T1]{fontenc}
\usepackage{verbatim}
\usepackage{amssymb,amsmath}
\usepackage{xcolor}
\usepackage{float}
\usepackage{cite}
 \usepackage{hyperref}
\hypersetup{
     colorlinks=true,
     linkcolor = red,
     anchorcolor = blue,
     citecolor = blue,
     filecolor = blue,
     urlcolor = blue
     }

\def\p{\partial}

\allowdisplaybreaks

\def\nn{\nonumber}

\def\be{\begin{eqnarray}}
\def\ee{\end{eqnarray}}

\newcommand\blfootnote[1]{%
  \begingroup
  \renewcommand\thefootnote{}\footnote{#1}%
  \addtocounter{footnote}{-1}%
  \endgroup}

\numberwithin{equation}{section}

\begin{document}


\begin{titlepage}
	\thispagestyle{empty}
	\begin{flushright}

	\end{flushright}
\vspace{35pt}

	\begin{center}
	    { \Large{
	    Three-forms and Fayet--Iliopoulos terms in Supergravity: 
	    \\[0.3cm]
	    Scanning Planck mass and BPS domain walls 
	    }}

		\vspace{50pt}

		{Niccol\`o Cribiori$^1$, Fotis Farakos$^2$ and George Tringas$^{3}$}

		\vspace{25pt}

\begin{center}

\vspace{0.5cm}
\textit{\small $^1$  Institute for Theoretical Physics, TU Wien, Wiedner Hauptstrasse 8-10/136, A-1040 Vienna, Austria 
\\[0.2cm]
$^2$ KU Leuven, Institute for Theoretical Physics,  Celestijnenlaan 200D, B-3001 Leuven, Belgium 
\\[0.2cm]
$^3$ Physics Division, National Technical University of Athens 15780 Zografou Campus, Athens, Greece } 
\blfootnote{e-mails: \href{mailto:niccolo.cribiori@tuwien.ac.at}{niccolo.cribiori@tuwien.ac.at}, \href{mailto:fotios.farakos@kuleuven.be}{fotios.farakos@kuleuven.be},
\href{mailto:georgiostringas@mail.ntua.gr}{georgiostringas@mail.ntua.gr}}
\end{center}

		\vspace{40pt}

		{ABSTRACT}
	\end{center}

	\vspace{10pt}

We embed a new three-form vector multiplet in ${\cal N}=1$ supergravity and we show that it can be used to generate dynamically the Hilbert--Einstein term. We then recast the theory into the standard Freedman model and we argue that a pure Fayet--Iliopoulos term is in tension with the Weak Gravity Conjecture. Finally, we couple the three-form to a super-membrane and study BPS domain walls within matter-coupled supergravity. In these models, the Planck mass takes different values on the domain wall sides.

\vfill

\end{titlepage}



{\hypersetup{hidelinks}
\tableofcontents
}

\setcounter{footnote}{0}

\baselineskip 5.6 mm

\section{Introduction}

Gauge three-forms in four dimensions have been studied extensively, since they can induce interesting physical effects despite the fact that they cannot propagate any physical degrees of freedom. One of the first applications was in cosmology, where they have been employed to construct models in which the cosmological constant was generated dynamically \cite{Brown:1987dd,Bousso:2000xa,Feng:2000if}. More recently they have been studied, for example, in inflation \cite{Kaloper:2008fb,Germani:2009iq,Germani:2009gg,Marchesano:2014mla,Dudas:2014pva,Lee:2019aci,Almeida:2019xzt} and in relation to naturalness problems \cite{Dvali:2005an, Dvali:2005zk, Giudice:2019iwl,Kaloper:2019xfj,Lee:2019efp,Lee:2019twi,Bordin:2019fek}. Three-forms have been also embedded into supersymmetric models, both in global supersymmetry \cite{Gates:1980ay,Gates:1980az,Nishino:2009zz,Bandos:2010yy,Groh:2012tf,Nishino:2013oea,Cribiori:2018jjh,Nitta:2018vyc,Nitta:2018yzb,Bandos:2019qok} and in supergravity \cite{Binetruy:1996xw,Ovrut:1997ur,Bandos:2011fw,Farakos:2016hly,Bandos:2016xyu,Aoki:2016rfz,Buchbinder:2017vnb,Kuzenko:2017vil,Bandos:2019lps,Dudas:2019gxd}. In this context, they can be used for instance to dynamically introduce (some of) the parameters which are present in effective theories coming from flux compactifications of string theory 
to four dimensions \cite{Bielleman:2015ina,Carta:2016ynn,Valenzuela:2016yny,Farakos:2017jme,Becker:2017zwe,Bandos:2018gjp,Herraez:2018vae,Lanza:2019xxg}.

In this work, we continue the study of gauge three-forms in theories with local supersymmetry. We start by constructing a new, composite abelian vector multiplet in minimal supergravity, in which the real auxiliary field D is replaced by a composite expression, containing the Hodge dual of the gauge three-form. This multiplet can be viewed as an extension to local supersymmetry of an analogous three-form vector multiplet, which was studied in the global case in \cite{Gates:1983nr, Antoniadis:2017jsk,Cribiori:2018jjh}. Then, we couple the new vector multiplet to gravity and we propose a manifestly supersymmetric Lagrangian describing its interactions. This Lagrangian is one of the main results of the work and it presents peculiar features. As we will show, in fact, the off-shell form of the Lagrangian is remarkably different from standard supergravity coupled to an abelian vector multiplet. In particular, as we will see, the differences do not end in substituting solely the auxiliary field with the gauge three-form, as it is the case in global supersymmetry \cite{Cribiori:2018jjh}. Eventually, when going on-shell, the known supergravity Lagrangian is recovered, together with an important novelty: the (reduced) Planck mass $M_P$ is generated dynamically and given by 
\be
\label{MP2}
M_P^2 = n \, , 
\ee
where the mass dimension two constant $n$ is associated to the three-form flux and has to be non-negative for a consistent theory of gravity.

To understand why this feature can be considered as new with respect to known constructions, let us recall the logic behind the so-called superconformal and super-Weyl invariant formulations of supergravity, which are among the most employed ones. In these approaches, one usually starts from an action which is invariant under local superconformal or super-Weyl symmetry, of which local supersymmetry is a subgroup. In this way, the couplings are highly constrained by the large symmetry group from the very beginning. Then, in order to obtain a theory of gravity, superconformal or super-Weyl invariance have to be broken to local supersymmetry and the Planck scale has to be introduced into the theory. This step is usually accomplished with the help of a compensator field, which is appropriately fixed by a field redefinition. For example, in the superconformal approach of \cite{Freedman:2012zz} to minimal supergravity, the compensator is a chiral multiplet $\phi$, which is eventually fixed at $\phi= M_P$. In general, different off-shell formulations of supergravity are related to different choices of the compensator multiplet. 

Our construction, instead, follows an alternative logic: it does not require the introduction of a specific compensator multiplet and the Planck mass is generated dynamically, once the equations of motion for the gauge three-form are solved. Indeed, the superspace Lagrangian that we propose for the new vector multiplet is super-Weyl invariant off-shell, since no physical scale is present and despite the fact that we do not have any additional (compensator) superfield, besides the three-form vector superfield. As we will show, super-Weyl invariance is then spontaneously broken, when integrating out the gauge three-form. In fact, the equations of motion set the Hodge dual of the gauge three-form to be a constant, with mass dimension two. This is precisely the Newton constant, or Planck mass, and gravity becomes dynamical on-shell.

Notice that, within the standard superconformal approach, one might try to introduce dynamically the Newton constant, by giving a scalar potential to the compensator and by implementing a Higgs mechanism. This strategy is different from ours and, moreover, it would require particular attention, since the compensator usually has the wrong sign in its kinetic term.

After recovering the on-shell supergravity action for a vector multiplet, we take a step forward and we show that our construction can be dualized into a more familiar model. This is the Fayet--Iliopoulos model with D-term breaking of supersymmetry, whose supergravity embedding has been proposed by Freedman \cite{Freedman:1976uk}. We take this opportunity to comment on the fact that pure Fayet--Iliopoulos terms in supergravity are in tension with the Weak Gravity Conjecture \cite{ArkaniHamed:2006dz}. This can be understood as an extension of previous no-go theorems on Fayet--Iliopoulos terms in generic theories of quantum gravity \cite{Komargodski:2009pc}, which would place them within the Swampland \cite{Palti:2019pca}.

Finally, we discuss the couplings of the system to matter chiral superfields and to super-membranes. In this respect, we give two different examples. The first one involves only a single charged chiral superfield, besides the three-form vector superfield. When adding a membrane to the setup and trying to construct 1/2-BPS domain wall solutions interpolating between the distinct supersymmetric AdS vacua on the two sides, we notice that the profile has various irregularities and we point out the root of the problem. As a consequence, we do not pursue with the analysis further. Instead, we give a second example where higher derivative terms are introduced, which however can be dualized into standard gauged supergravity coupled to two massive chiral superfields. Once a membrane is added to the system, we construct and discuss smooth domain wall solutions between the two different AdS regions on each of its sides.

The models in sections \ref{sec:singdw} and \ref{sec:smoothdw} are constructed by coupling to matter the original model in section \ref{sec:puremod}. The first of these models is described by a standard, two-derivatives gauged supergravity action, eventually coupled to membranes, but it leads to an irregular domain wall profile. In section \ref{sec:smoothdw}, we cure this problem by letting the auxiliary field of the supergravity multiplet propagate. This strategy leads to a regular domain wall solution, at the cost of introducing higher derivatives. However, these are avoided in our presentation by means of an appropriate dualization to a standard, two-derivatives, gauged supergravity. For these reasons, the models in sections  \ref{sec:puremod},  \ref{sec:singdw} and \ref{sec:smoothdw} are clearly different from each other, nevertheless it is instructive to present them in the proposed order of increasing complexity.

We stress that the supersymmetric AdS vacua that we construct and that are separated by a membrane are characterized by a different value of the flux parameter $n$, which normalizes the Hilbert--Einstein term. When crossing the membrane, this parameter jumps as 
\be
\Delta n \sim Q \, , 
\ee
$Q$ being the membrane tension. In this sense, we see that in our construction \emph{the Planck mass $M_P$ is a variable that scans}.

In this work we adopt the old minimal formulation of supergravity, following the conventions of \cite{Wess:1992cp}.

\section{Three-form gauge multiplet in supersymmetry}

In this section, we review the embedding of a gauge three-form into an abelian vector superfield in global supersymmetry. This discussion is meant to be a warm-up for the next section, where we are going to generalize and extend the results to supergravity. Since a complex linear superfield is a central ingredient in the construction, in supersymmetry as well as in supergravity, we start by reviewing its main properties.

In global supersymmetry, a complex linear superfield $\Sigma$ is defined by the superspace condition 
\be
\overline D^2 \Sigma =0,
\ee
which has solution
\begin{equation}
\begin{split}
\Sigma = &\sigma + \sqrt 2\theta \varphi + \sqrt{2} \overline\theta \overline\chi - \theta \sigma_m \overline\theta {P}^{m} + \theta^2 \overline s+ \sqrt 2 \theta^2\overline\theta \overline\zeta\\
&-\frac{i}{\sqrt{2}} \overline\theta^2 \theta \sigma^m \partial_m \overline\chi + \theta^2\overline\theta^2 \left(\frac{i}{2} \partial_m {P}^{m} -\frac14 \Box \sigma \right),
\end{split}
\label{CompS}
\end{equation}
where $\sigma$ and $s$ are complex scalar fields,  $\varphi$, $ \chi$ and $\zeta$ are Weyl fermions and  ${P}^m$ is a complex vector. Alternatively, we can define the components of $\Sigma$ by using superspace derivatives, as is more convenient in supergravity. For example, we have
\begin{equation}
\frac14 \bar{\sigma}^{m\,\dot\alpha \alpha}
\left[D_\alpha,\overline{D}_{\dot{\alpha}}\right] \Sigma| = {P}^m \equiv- v^m + i C^m\,,
\end{equation}
where $v^m =- {\rm Re}\, {P}^m$ is a real vector and $C^m = {\rm Im}\, {P}^m$ can be interpreted as the Hodge dual of a three-form $C_{mnp}$
\begin{equation}
\label{Csusy}
C^m =\frac{1}{3!}\varepsilon^{mnpq} C_{npq}= \frac14 \bar{\sigma}^{m\,\dot\alpha \alpha}
\left[D_\alpha,\overline{D}_{\dot{\alpha}}\right] {\rm Im}\,\Sigma|\,. 
\end{equation}
From $\Sigma$, we can define a composite vector superfield U as
\be
U = -\frac{\Sigma + \overline \Sigma}{2}.
\ee
This superfield was first introduced in \cite{Gates:1983nr} and then studied further in \cite{Antoniadis:2017jsk,Cribiori:2018jjh}. Here, we mainly follow the analysis of \cite{Cribiori:2018jjh}.

Given the properties of $\Sigma$, one can easily check that $U$ is a well-defined vector superfield. Indeed, it contains a vector component field 
\be
\frac14 \bar{\sigma}^{m\,\dot\alpha \alpha}\left[D_\alpha,\overline{D}_{\dot{\alpha}}\right] U| &=  v^m
\ee
and, under the shift
\be
\label{SY}
\Sigma \to \Sigma + 2Y,
\ee
where $Y$ is chiral superfield, it transforms as a gauge vector superfield
\be
U \to U + Y + \overline Y.
\ee
Furthermore, under 
\be
\label{SL}
\Sigma \to \Sigma - i L,
\ee
where $L$ is a real linear superfield, namely a complex linear superfield subject also to a reality constraint, $L = L^*$, the three-form transforms as
\be
C^m \to C^m +\frac{1}{3!}\epsilon^{mnpq}\partial_n B_{pq},
\ee
with $B_{pq}$ given by
\be
-\frac14 \overline \sigma^{m\, \dot\alpha \alpha}[D_\alpha,\overline D_{\dot\alpha}]L| = \frac{1}{3!}\epsilon^{mnpq}\partial_n B_{pq}.
\ee
As a consequence, we can interpret $C_{mnp}$ as a gauge three-form, since it has the proper gauge transformation. Notice also that $U$ is invariant under \eqref{SL}.

Given a vector superfield $U$, one can define a gauge-invariant field strength chiral superfield $\mathcal{W}_\alpha$ as
\be
\mathcal{W}_\alpha (U)= -\frac14 \overline D^2 D_\alpha U.
\ee
Then, a supersymmetric Lagrangian for $U$ is
\be
\label{Lsusy}
\mathcal{L}=\frac{1}{4 g^2}\left( \int {\rm d}^2 \theta\, \mathcal{W}^2(U) + c.c.\right) + \mathcal{L}_{bd},
\ee
where 
\be
\begin{aligned}
\mathcal{L}_{bd} &= -\frac{1}{16 g^2}\int {\rm d}^4 \theta D^\alpha \mathcal{W}_\alpha (\Sigma - \overline \Sigma) + c.c.\\
&=\frac{1}{64 g^2}[D^2,\overline D^2](D^\alpha \mathcal{W}_\alpha (\Sigma-\overline \Sigma) )|
\end{aligned}
\ee
is a boundary term contribution, which is necessary to take a consistent variation of the gauge three-form. Indeed, as we will see below, the gauge three-form is contained within the D-term component field of $U$, which is auxiliary. The standard procedure for eliminating the auxiliary fields and going on-shell consists in solving the corresponding equations of motion and plugging them back into the Lagrangian, in order to generate a scalar potential. Since the gauge three-form is auxiliary in our construction, it has to be integrated out as any ordinary auxiliary field. However, without the proper boundary term, we would obtain a wrong sign in the scalar potential, once the solution of the equations of motion is inserted back into the Lagrangian. More details about this can be found in \cite{Cribiori:2018jjh}, together with a systematic procedure to compute explicitly $ \mathcal{L}_{bd}$ for the three-form vector multiplet. In the next section, we are going to generalize to supergravity also such a discussion concerning boundary terms. 

The bosonic sector of \eqref{Lsusy} is 
\be
\mathcal{L} = -\frac{1}{4g^2} F_{mn} F^{mn} + \frac12 {\rm D}^2 + \mathcal{L}_{bd}, \qquad \mathcal{L}_{bd} = -\frac{1}{g^2}\partial_m ({\rm D}C^m),
\ee
where D is the real auxiliary field of the vector superfield, D$=-\frac12 D^\alpha \mathcal{W}_\alpha|$. Since $U$ is composite, its auxiliary field D is also a composite expression and it is given in terms of the three-form as
\be
\label{Dsusy}
{\rm D} = -\frac12 D^\alpha \mathcal{W}_\alpha(U)| = \partial^m C_m = *{\rm d}C_3 \, .
\ee
Therefore, in this construction the auxiliary field D is replaced by the Hodge dual of the gauge three-form, as anticipated above. We will see that, in supergravity, the generalization of \eqref{Dsusy} will contain interesting and important differences. In particular, besides the gauge three-form, the composite auxiliary field D is going to contain also fields of the gravity multiplet.

In \cite{Cribiori:2018jjh}, it is also shown how to recast \eqref{Lsusy} into a dual theory, which is the Fayet--Iliopoulos model for D-term supersymmetry breaking. In particular, the Fayet--Iliopoulos parameter is generated dynamically, once the three-form is integrated out. In the next section, we are going to describe also the generalization of such a dualization procedure in supergravity.

\section{Three-form gauge multiplet in supergravity} 
\label{sec:puremod}
In this section, we first introduce a new vector multiplet in supergravity which contains a three-form inside its D-term component field. Then, we propose a Lagrangian in superspace which describes its interactions. In an appropriate Wess--Zumino (WZ) gauge, such a Lagrangian has a classical scale invariance, which is broken when the three-form is integrated out and a physical scale is inserted dynamically into the theory. With a dualization procedure that we are going to discuss, this theory can be recast into the more familiar Fayet--Iliopoulos model for D-term breaking in supergravity. Finally, we comment on the tension between pure Fayet--Ilioupoulos terms and the Weak Gravity Conjecture.

\subsection{A new vector multiplet with a three-form} 
Given a complex linear superfield $\Sigma$, such that
\be
\label{CLS}
- \frac14 \left( \overline{\cal D}^2 - 8 {\cal R} \right) \Sigma = 0 \,,
\ee
we define a new, composite vector superfield $U$ as
\be
\label{VeryNice}
U = - \frac12 \log \left( \Sigma + \overline \Sigma \right) \,. 
\ee
This superfield is manifestly real and it is the supergravity generalization of the three-form vector superfield discussed in the previous section in global supersymmetry.\footnote{We would like to thank Stefano Lanza for discussions related to this multiplet, while the work \cite{Cribiori:2018jjh} was under completion.}

Notice that $U$ is left invariant under the shift 
\begin{equation}
\label{Lgauge}
    \Sigma \to \Sigma -i L,
\end{equation}
where $L$ is a real linear superfield. Keeping this in mind, it is instructive to study the behavior of $U$ under super-Weyl transformations. In our conventions, these are defined to act on the super-vielbein as
\begin{align} 
\label{weylmetric}
\delta {E_M}^a &= (Y+\overline Y) {E_M}^a,\\
\label{weylgravitino}
\delta {E_M}^\alpha &= (2\overline Y - Y){E_M}^\alpha+\frac{i}{2} {E_M}^b{(\epsilon \sigma_b)^\alpha}_{\dot \alpha}\overline{\mathcal D}^{\dot\alpha}\bar Y,
\end{align}
where $Y$ is a chiral superfield. The super-Weyl transformations of a real linear superfield $L$ and of a complex linear superfield $\Sigma$ are given respectively by (see for example \cite{Kuzenko:2017vil})
\begin{align}
\label{sWeylL}
   L&\to e^{-2 Y-2\overline Y} L,\\ 
   \label{sWeylSigma}
   \Sigma&\to e^{w Y-2\overline Y} \Sigma.  
\end{align}
In general, the weight $w$ of the complex linear superfield is arbitrary. However, if we want to preserve the symmetry \eqref{Lgauge} of $U$, we are forced to choose $w=-2$. Indeed, in this way, under a super-Weyl transformation of $\Sigma$, the composite superfield $U$ transforms as
\be
\label{UtoU}
U \to U + Y + \overline Y \,  
\ee
and the transformed superfield still enjoys \eqref{Lgauge}. The transformation \eqref{UtoU} is the superspace gauge transformation of a vector superfield, which induces the standard gauge variation on the vector component field 
\begin{equation}
\label{UR1}
    v_m \to v_m + \partial_m a, \qquad a = 2{\rm Im} (Y)|.
\end{equation}
The embedding of the component fields inside the composite superfield $U$ will be discussed below.

To summarize, the requirement that the composite superfield $U$ is invariant under the shift \eqref{Lgauge} fixes the form of the variation of $U$ under super-Weyl transformations to be that of a gauge transformation \eqref{UtoU}. As we will discuss further in the following, this fact implies that the R-symmetry is gauged in our setup. 
Indeed, the gravitino, which is always charged under the R-symmetry, 
is charged under the super-Weyl transformation \eqref{weylgravitino} and transforms under \eqref{UR1} as 
\begin{equation}
    \psi_m \to  e^{-3 i {\rm Im} Y|}\psi_m \, , 
\end{equation}
taking into account, in \eqref{weylgravitino}, that $(2 \overline Y - Y) |_{a} = -\frac32 i a$.\footnote{Here $|_{a}$ means that we look only at the `a' transformation, namely we set Re$Y=0$.} Another way to see that the R-symmetry is gauged is by recalling that the supergravity multiplet contains a real superfield $G_{\alpha\dot\alpha}$, which has a vector auxiliary field in the lowest component, $G_{\alpha\dot\alpha}|=-\frac13 b_{\alpha \dot\alpha}=-\frac13 \sigma^c_{\alpha \dot \alpha}b_c$. Under a super-Weyl transformation, the superfield $G_{\alpha\dot\alpha}$ transforms as
\begin{equation}
    \delta G_{\alpha \dot\alpha}=-(Y+\overline Y)G_{\alpha\dot\alpha} + {i}\mathcal{D}_{\alpha\dot\alpha}(\overline Y-Y).
\end{equation}
Therefore, if we consider only the part of the super-Weyl transformations generated by $a=2 {\rm Im}(Y)|$, we can see directly that $b_a$ transforms as a gauge vector, namely
\begin{equation}
\label{U1Rb}
     b_c \to b_c - 3 e^m_c\partial_m a. 
\end{equation}
This is yet another signal of the gauging of the R-symmetry.

We proceed now and study the component structure of the composite superfield $U$, while we will present a supergravity Lagrangian in the next subsection. By taking advantage of all of the freedom that is granted by the combined super-Weyl and shift transformations
\be
\label{GandSW}
\Sigma \to e^{-2 Y - 2 \overline Y} \left(\Sigma - {i} L \right)  \, ,
\ee
which is the supergravity generalization of the combination of \eqref{SY} and \eqref{SL}, we can write the multiplet $U$ in an appropriate WZ gauge. We will restrict our attention mainly to the bosonic sector. We start by giving the independent bosonic component fields of the complex linear superfield $\Sigma$ contained in $U$, which are 
\be
\begin{aligned}
\label{sigmacomp}
\Sigma| & = \sigma \, , 
\\ 
-\frac14 {\cal D}^2 \Sigma| & = F \, , 
\\
\overline{\cal D}_{\dot{\alpha}} {\cal D}_{\alpha}\Sigma|&= P_{\alpha \dot \alpha}\, ,
\end{aligned}
\ee
where the fields $\sigma$, $F$ and $P_{\alpha \dot\alpha}$ are complex. Then, we notice that the shift symmetry \eqref{Lgauge} allows us to set
\be
{\text Im} \, \sigma \Big{|}_{WZ} = 0,  
\ee 
by means of an appropriate gauge choice for the lowest component real scalar of $L$. If the theory is gauge invariant, we can also use the gauge transformation \eqref{UtoU}, or equivalently \eqref{sWeylSigma} with $w=-2$, in order to fix
\begin{equation}
\label{WZ1}
{\text Re} \, \sigma \Big{|}_{WZ} = \frac12, \,  \quad  \text{i.e.} \quad U \Big{|}_{WZ} = 0.
\end{equation}
In this gauge, we can set 
\begin{equation}
\qquad\, {\cal D}^2 U \Big{|}_{WZ} = 0, \, \, \quad  \text{i.e.} \quad  F\Big{|}_{WZ} = - \frac16 \overline M \, , 
\end{equation}
where the complex scalar $M$ is an auxiliary field of the supergravity multiplet. 
Among the non-gauge degrees of freedom, inside $U$ we have a vector component field, defined as
\be
\label{vdef}
v_{\alpha \dot \alpha} = -\frac12 [{\cal D}_\alpha , \overline{\cal D}_{\dot \alpha}] U \Big{|} = -\text{Re}P_{\alpha \dot{\alpha}}\Big{|}_{WZ} \, ,
\ee
which has been already introduced and which transforms as \eqref{UR1}. Notice that the $P_{\alpha \dot \alpha}$ appearing in \eqref{vdef} is the WZ gauge-fixed version of the quantity in \eqref{sigmacomp}.\footnote{We recall that in the WZ gauge we have
\be
P_{\alpha \dot \alpha}\Big{|}_{WZ} = \overline{\mathcal{D}}_{\dot \alpha}\mathcal{D}_\alpha\Sigma\Big{|}_{WZ}=-\frac12[\mathcal{D}_\alpha, \overline{\mathcal{D}}_{\dot\alpha}]\Sigma\Big{|}_{WZ}.
\ee
In particular, torsion terms are vanishing, $T^C_{\alpha\dot\alpha}\mathcal{D}_C \Sigma\Big{|}_{WZ}=0$.}  We can now define the Hodge dual of the three-form gauge field 
\be 
C^m = \frac{1}{3!}\epsilon^{mnpq}C_{npq} \, , 
\ee
as \cite{Ovrut:1997ur} 
\be
C^m =\frac14 \left(\overline \sigma^{m\, \dot \alpha \alpha}[\mathcal{D}_\alpha, \overline{\mathcal{D}}_{\dot\alpha}]+8 G^m\right){\rm Im}\Sigma\Big{|}.
\ee
This is the generalization of \eqref{Csusy} to supergravity. In the WZ gauge, it reduces to
\be
C^m =\frac14 \overline \sigma^{m\, \dot \alpha \alpha}[\mathcal{D}_\alpha, \overline{\mathcal{D}}_{\dot\alpha}]{\rm Im}\Sigma\Big{|}_{WZ} = {\rm Im} P^m\Big{|}_{WZ}.
\ee
Notice that this field transforms properly under the gauge transformation \eqref{Lgauge}, namely
\be
\label{Cgauge}
C_{mnp} \to C_{mnp} + \p_{[m} B_{np]},
\ee
where $B_{np}$ is given by
\be
-\frac14 \left(\overline \sigma^{m\, \dot \alpha \alpha}[\mathcal{D}_\alpha, \overline{\mathcal{D}}_{\dot\alpha}]+8 G^m\right) L\Big{|} = \frac{1}{3!} \epsilon^{mnpq}\partial_n B_{pq}. 
\ee
Indeed, one can think of the transformation \eqref{Cgauge} as a defining property for $C_{mnp}$, which is indicating that such a quantity is a gauge three-form. We stress that this interpretation is consistent only within actions that are invariant under \eqref{Lgauge}.

As usual, given a vector superfield $U$ one can define a gauge invariant field strength superfield $\mathcal{W}_\alpha$ as
\be
{\cal W}_\alpha = - \frac14 \left( \overline{\cal D}^2 - 8 {\cal R} \right) {\cal D}_\alpha U \, .
\ee
This superfield is chiral and transforms under (\ref{GandSW}) as 
\be
\label{WUW} 
{\cal W}_\alpha \rightarrow {\cal W}_\alpha \, e^{-3 Y} \, . 
\ee
Its independent fermionic component field is the gaugino $\lambda_\alpha$ in the lowest component
\be
{\cal W}_\alpha | = -{i} \lambda_\alpha \, , 
\ee
while for what concerns the independent bosonic components, we have the field strength $F_{mn}$, encoded into 
\be
({\cal D}_\alpha W_\beta + {\cal D}_\beta W_\alpha) | =  -4{i} (\sigma^{ba} \epsilon)_{\alpha \beta} \hat D_b v_a 
= 2{i} (\sigma^{ba} \epsilon)_{\alpha \beta} \hat F_{ab},
\ee
where we defined the supercovariant derivative
\begin{equation}
\hat D_b v_{\alpha \dot \alpha} = e_{b}^{m} \left\{ D_m v_{\alpha \dot \alpha} 
+ {i} (\psi_{m \alpha} \overline \lambda_{\dot \alpha} + \overline \psi_{m \dot \alpha} \lambda_\alpha ) 
+ \frac{i}{2} \psi_m v \overline \psi_a \sigma^a_{\alpha \dot \alpha} 
\right\} \, , 
\end{equation}
and $D_m$ is the covariant derivative which includes the spin-connection $\omega_{ma}^{\ \ \ b}(e,\psi)$. 

A vector superfield has also a D-term component field D, which is independent but (usually) does not propagate physical degrees of freedom.  However, in our case, since we are considering the composite vector superfield $U$ given in \eqref{VeryNice}, the D-term component field D is also composite and given by 
\be
\label{Dcomp}
\text{D} = -\frac12 {\cal D}^\alpha {\cal W}_\alpha | =  \frac{1}{6}R+  
\nabla_m C^m
-\frac{1}{9}(b_m+3v_m)^2  +\frac19 M \overline M+ \text{fermions}. 
\ee
This generalizes in a non-trivial way the simple result \eqref{Dsusy}, valid in global supersymmetry. In the following, we will often use the abbreviation $\nabla_m C^m\equiv * {\rm d}C_3$.

At this point, it is worth commenting on the following facts. 
First, notice that the expression \eqref{Dcomp} contains explicitly the Ricci scalar $R$. However, even when multiplied by the determinant $e$, \eqref{Dcomp} is not transforming to a total derivative under local supersymmetry, therefore it cannot provide a supersymmetric Lagrangian density. Second, we recall that the D-term component of a vector superfield is gauge-invariant. Since the expression \eqref{Dcomp} contains the combination $(b_m+3v_m)$ and since $v_m$ transforms as \eqref{UR1} under gauge transformations, then also $b_m$ has to transform, in order for such a combination to be gauge-invariant. The correct transformation of $b_m$, which cancels that of $v_m$, is given precisely by \eqref{U1Rb}. This is a signal of the gauging of the R-symmetry, at least in a supersymmetric theory containing a linear term in D.

Before concluding this presentation, let us discuss also what happens in the trivial case in which we set the gauge superfield $U$ to be pure gauge. A direct way to obtain this is by choosing
\be
\label{puregauge}
U \equiv 0 \quad \to \quad \Sigma = \frac12 \, . 
\ee
Notice that this is different from the WZ gauge-fixing condition \eqref{WZ1}. Indeed, we are now setting the entire superfield $U$ to zero, without breaking supersymmetry. Then, when applied to \eqref{CLS}, the requirement \eqref{puregauge} would also imply that 
\be 
{\cal R} = 0 \, . 
\ee
This equation is nothing but the supersymmetrization of the Einstein equations without sources. As a result, we see that setting the gauge multiplet to be pure gauge forces also to impose the full Einstein equations for the supergravity multiplet. An alternative way to obtain the same result is to set $U \equiv \Phi + \overline \Phi$, where $\Phi$ is a chiral superfield. Indeed, this is the more generic form of a pure gauge vector superfield. Then, we would find that $\mathcal{W}_\alpha(U) \equiv 0$, which in turn would give a vanishing D-term auxiliary field, namely D$\equiv 0$. Applying this result to \eqref{Dcomp}, we would see that it implies again the Einstein equations without sources, namely $R=0$ and $M=0$ in components. In fact, since $M$ is vanishing, the full multiplet $\mathcal{R}$, which has $M$ as lowest component, vanishes as well due to supersymmetry. Notice also that this poses restrictions on the superfield $G_a$, which is related to $\mathcal{R}$ by the old-minimal supergravity Bianchi identities, namely $4 {i} {\cal D}_a G^a = {\cal D}^2 {\cal R} - \overline{\cal D}^2 \overline{\cal R} \equiv  0$ and $\overline{\cal D}^{\dot \alpha} G_{\alpha \dot \alpha} = {\cal D}_\alpha {\cal R} \equiv 0$. A similar effect occurs also in the new-minimal formulation of supergravity, which is described by a gauge multiplet $V_{\rm R}$. In particular, one can see that setting $V_{\rm R}$ to be pure gauge leads directly to the Einstein equations without sources.

\subsection{The supergravity action and dynamical Planck mass} 

In this section we present a supergravity Lagrangian for the new, composite vector multiplet $U$ and we discuss its main properties. First, we work at the component fields level and we show how the integration of the gauge three-form generates dynamically the four-dimensional Planck mass. Then, we perform a complementary analysis in superspace and we recast the theory into its dual formulation without the three-form, proving the equivalence with the Freedman's model for the Fayet--Iliopoulos D-term in supergravity \cite{Freedman:1976uk}. Finally, we argue that pure Fayet--Iliopoulos terms are generically incompatible with the Weak Gravity Conjecture.

\subsubsection{Lagrangian and component fields analysis}

We want now to construct a simple Lagrangian which describes the coupling of the new three-form gauge vector multiplet in supergravity. In particular, such a Lagrangian has to be invariant under the transformations \eqref{GandSW}, for consistency. Let us first recall that, under these transformations, the real and chiral superspace densities change respectively as 
\be
\label{SWI}
{\rm d}^4 \theta \, E \rightarrow {\rm d}^4 \theta \, E \, e^{2 Y + 2 \overline Y} 
\, , \quad  
{\rm d}^2 \Theta \,  2{\cal E} \rightarrow {\rm d}^2 \Theta \,  2{\cal E} \, e^{6 Y} \, , 
\ee
while the chiral projector transforms as 
\be
\label{WPR} 
- \frac14 \left( \overline{\cal D}^2 - 8 {\cal R} \right) \to
- \frac14 \left( \overline{\cal D}^2 - 8 {\cal R} \right) e^{-4 Y + 2 \overline Y} \, . 
\ee 
Taking these transformations into account, a simple Lagrangian which is invariant under \eqref{GandSW} is then
\be
\label{LG1}
{\cal L} = \left( \frac{1}{4g^2} \int {\rm d}^2 \Theta \, 2 {\cal E} \, {\cal W}^2(U) + c.c.  \right)  
+ {\cal L}_{bd} \, ,  
\ee
where ${\cal L}_{bd}$ is a boundary term that is required for a consistent variation of the action, as explained previously, and is given by 
\be
\label{BBDDD}
{\cal L}_{bd} =  - \frac{1}{16 g^2}  \int {\rm d}^2 \Theta \, 2 {\cal E} \left( \overline{\cal D}^2 - 8 {\cal R} \right) 
\Big{[} {\cal D}^\alpha{\cal W}_\alpha \, 
\frac{\Sigma - \overline \Sigma}{\Sigma + \overline \Sigma} \Big{]} + c.c. 
\ee
As we will see, despite the fact that there is only one single superspace term, besides the boundary contribution, this Lagrangian contains both the kinetic terms for the gravity multiplet and for the vector multiplet. Moreover, here we are giving directly the form of the boundary term without a derivation, but later we will describe the procedure that explicitly gives rise to \eqref{BBDDD}.

Before studying further the Lagrangian \eqref{LG1}, it is instructive to discuss the reasons why other proposals, which in principle could appear simpler, would actually be problematic. Arguably, the simplest construction invariant under the symmetries \eqref{GandSW} would be a term depending on the real or on the imaginary part of $\Sigma$, namely
\be
\label{nope}
\int {\rm d}^4 \theta \, E \,  (\Sigma+\overline \Sigma) \, \qquad \text{or}\qquad { i}\int {\rm d}^4 \theta \, E \,  (\Sigma-\overline \Sigma)\,. 
\ee
When expanding in components, in the WZ gauge we will find that 
\begin{align}
\label{Lfailed1}
    \int {\rm d}^4 \theta \, E \,  (\Sigma+\overline \Sigma) &= - e \,\nabla_m C^m + \text{fermions},\\
    \label{Lfailed2}
    { i}\int {\rm d}^4 \theta \, E \,  (\Sigma-\overline \Sigma) &= \frac e3 e^m_a \mathcal{D}_m b^a+ e \,\nabla_m v^m + \text{fermions}.
\end{align}
In both cases, the component expansion does not contain the Ricci scalar. Therefore, we can readily conclude that despite their simplicity, \eqref{Lfailed1} and \eqref{Lfailed2} are not good proposals for a supergravity Lagrangian.

Having motivated our choice of the Lagrangian \eqref{LG1}, we can now proceed and study what is the supergravity theory described by it. For simplicitly, we will first work at the component level, in order to present the main physical properties more directly. Then, in section \ref{sec:dualform}, we will also perform a complementary analysis in superspace, in order to capture the full dynamics of the theory. Once expanded in component fields, the bosonic sector of \eqref{LG1} in the WZ gauge is
\be
\label{ActionBefore}
e^{-1} {\cal L} = - \frac{1}{4g^2} F_{mn} F^{mn} 
+ \frac{1}{2 g^2} \left( \frac{1}{6}R+ *\text{d}C_3-\frac{1}{9}(b_m+3v_m)^2 +\frac19 M\overline M \right)^2  
+ {\cal L}_{bd}  \, , 
\ee 
where the (bosonic sector of the) boundary term is
\be
\begin{aligned}
\label{Lbdcomp}
e^{-1}{\cal L}_{bd} & = -\frac{1}{g^2} \mathcal{D}_a \Big( {\rm D}\, C^a\Big) 
+ \dots 
\\
& = -\frac{1}{g^2}\mathcal{D}_a \left[ \left(\frac{1}{6}R+ *\text{d}C_3-\frac{1}{9}(b_m+3v_m)^2+\frac19 M\overline M \right) C^a\right] 
+ \dots 
\end{aligned}
\ee
and dots stand for other boundary terms, which are not needed for varying consistently the gauge three-form. We prefer to postpone the derivation of the boundary term \eqref{Lbdcomp} for the time being, since we will show in section \ref{sec:dualform} how to construct it directly from superspace. However, already at this stage, we can convince ourselves that the boundary term has the correct form, since it allows to perform a variation of the three-form without imposing a non-gauge invariant boundary condition. Indeed, by varying the three-form, we find the equation of motion
\be
{\rm d\, D} = {\rm d}\left( \frac{1}{6}R+  *\text{d}C_3-\frac{1}{9}(b_m+3v_m)^2  +\frac19 M \overline M \right)=0,
\ee
which is solved by
\be
\label{sol*F4}
* \text{d}C_3 = -\frac{1}{6}R+\frac{1}{9}(b^m+3v^m)^2-\frac19 M\overline M -3 n g^2 \, , 
\ee
where $n$ is an integration constant with mass dimension 2, $[n]=2$. In particular, in order to cancel the variation $\delta_{C_3}\mathcal{L}_{bd}$, the gauge invariant boundary condition
\be
\delta (* {\rm d}C_3) \big{|}_{bd} = 0 \, 
\ee
has been used. This confirms that \eqref{Lbdcomp} is the correct boundary term. To obtain the complete on-shell theory, we have to integrate out also the auxiliary fields $M$ and $b_a$ of the gravity multiplet. The integration of $M$ is trivial and sets $M=0$. On the other hand, by integrating out $b_a$, one finds
\be
b_m = - 3v_m  \, , 
\ee
which is in accordance with the gauging of R-symmetry by the physical abelian gauge field $v_m$. Once we insert the on-shell values for the auxiliary fields and for the Hodge dual of the four-form flux \eqref{sol*F4}, we obtain eventually 
\be
\label{onshellbulk}
e^{-1}{\cal L} = - \frac{n}{2} R - \frac{1}{4g^2} F_{mn} F^{mn} - \frac{9}{2}g^2 n^2 \, . 
\ee
This is the on-shell bosonic sector of \eqref{LG1}, in the WZ gauge. It contains the kinetic terms for the graviton and the vector $v_m$, together with a constant positive scalar potential, $V = \frac{9}{2}g^2n^2$, generated by the gauging of the U$(1)$ R-symmetry. The fact that the scalar potential is positive is telling us that supersymmetry is spontaneously broken, the gaugino being the goldstino, and the gravitino acquires a mass as a consequence of the super-BEH mechanism.

At this stage, it is worth commenting on the physical implications of our findings. We have seen that the action \eqref{ActionBefore} contains a gauge three-form which, once integrated, gives rise to the standard Hilbert--Einstein term. In particular, the flux $n$ works as an effective Planck mass and gravity is dynamical. For a consistent propagation of gravity, the constant $n$ has to be positive. However, since a priori the flux $n$ is not fixed to be positive, one can also wonder what happens if it has a vanishing value. In this case, the only term that survives in the effective action \eqref{onshellbulk} is the vector kinetic term, $-\frac{e}{4g^2} \, F_{mn} F^{mn}$. Then, the variation of the U$(1)$ gauge vector leads to the standard Maxwell equations in curved space. On the other hand, the variation of the metric gives a constraint $T_{mn} = 0$, where $T_{mn}$ is the energy momentum tensor of the vector field. Such a setup describes a system where the metric is non-dynamical and the U$(1)$ gauge vector is constrained to have vanishing energy-momentum tensor. This observation indicates in fact that the action \eqref{ActionBefore} links two completely different systems characterized by $n=0$ and by $n > 0$. If, in addition, a membrane is coupled to the system described by \eqref{ActionBefore} and it is charged under the three-form gauge field, then a domain wall can interpolate between the two phases. We will construct and discuss models of this kind in sections \ref{sec:singdw} and \ref{sec:smoothdw}.

\subsubsection{Dynamical breaking of scale invariance}

The action \eqref{ActionBefore} has a classical scale invariance under 
\be
\label{scaleinv}
g_{mn} \to \Omega \, g_{mn} \, , \qquad v_m \to v_m \, , \qquad  C_{m} \to C_{m} \, , 
\ee 
where $\Omega$ is a positive constant. Clearly, such an invariance is a property of the full theory, including fermions, since there are no explicit scales entering the action. Its origin is rooted in the invariance of the total superspace action \eqref{LG1} under super-Weyl transformations, of which scale transformations are a subgroup. 

The WZ gauge fixing condition breaks the original super-Weyl invariance down to the scale invariance \eqref{scaleinv}. However, we would like to stress that this is a choice we made only for convenience of the presentation. One can in principle avoid any gauge-fixing and perform the calculation in its full generality. Then, the complete component Lagrangian would be super-Weyl invariant, as its superspace counterpart. In particular, in our setup, the role of the compensator is played by the (non-physical) chiral multiplet $\sigma$, living inside the vector multiplet $U$. This is different, for example, from the superconformal approach to supergravity \cite{Freedman:2012zz}, where the compensator is usually a separate chiral multiplet, with a wrong sign in the kinetic term.

The scale invariance \eqref{scaleinv} (or the original super-Weyl invariance if one does not perform any gauge fixing) is then spontaneously broken by the three-form flux $n$, when going on-shell. This is unavoidable, since a physical scale, namely the Planck mass, is introduced in the action and the theory becomes supergravity. 
Notice that we are in fact generating the Planck mass together with the super-BEH mechanism. 

In this minimal model, once the three-form flux is introduced, we can fix it to $n=M_P^2$. Alternatively, we can perform a Weyl rescaling on \eqref{onshellbulk},
\be
\label{rescal}
g_{mn} \to \frac{M_P^2}{n} g_{mn} \, , 
\ee
which is not anymore a symmetry of the theory, and the action takes the standard form 
\be
\label{onshellbulk-n=1}
e^{-1}{\cal L} = - \frac{M_P^2}{2} R - \frac{1}{4g^2} F_{mn} F^{mn} - \frac92 g^2 M_P^4 \, . 
\ee
Notice that, after the rescaling \eqref{rescal}, the three-form flux $n$ has completely dropped out. This happens because the simple setup under investigation, with solely one three-form vector multiplet, does not include any other physical scales. Later, once we introduce more scales and other matter multiplets, we will see that the three-form flux value $n$ will not drop out of the theory, even after a Weyl rescaling.

\subsubsection{Dual superspace formulation and equivalence with Fayet--Iliopoulos D-term}
\label{sec:dualform}

We showed that, after the dynamical breaking of scale invariance, our construction describes a supergravity theory with spontaneously broken supersymmetry and a positive cosmological constant. We perform now a complementary analysis of the model in superspace. In particular, we show that its dual formulation, with the complex linear superfield integrated out, corresponds to the Freedman model for the standard Fayet--Iliopoulos D-term in supergravity \cite{Freedman:1976uk}. 
At the same time, we give a prescription to calculate the explicit form of the boundary term \eqref{Lbdcomp}.

In order to construct the dual formulation of the theory \eqref{LG1}, we do not start by assuming that the vector multiplet $U$ has the specific form \eqref{VeryNice} from the very beginning, rather we would like to obtain this result on-shell, by means of a Lagrange multiplier. Therefore, we start from the following Lagrangian for a standard vector superfield $U$ 
\be
\label{Dual1}
{\cal L} =  \frac{1}{4g^2} \int {\rm d}^2 \Theta \, 2 {\cal E} \, {\cal W}^2  
 - \frac34  \int {\rm d}^2 \Theta \, 2 {\cal E} 
 \left( \overline{\cal D}^2 - 8 {\cal R} \right) \, 
 \Lambda \left( \Sigma - \frac12 e^{-2U} \right) + c.c. \,, 
\ee
where $\Sigma$ is a complex linear superfield and $\Lambda$ is a Lagrange multiplier real superfield. Moreover, notice that the mass dimensions of the various superfields are 
\be
[\Lambda] = 2 \, , \qquad [\Sigma]=0 \, , \qquad [U]=0 \, , 
\ee
and that, in order for \eqref{Dual1} to be gauge invariant, we have to perform a super-Weyl transformation together with a gauge transformation \eqref{UtoU}.

First, we show that the Lagrangian \eqref{Dual1} is on-shell equivalent to our original proposal \eqref{LG1}. In particular, starting from \eqref{Dual1}, we can easily derive also the explicit form of the boundary term \eqref{BBDDD}. By varying \eqref{Dual1} with respect to $\Lambda$ and $U$, we get
\begin{align}
    \delta \Lambda: \qquad & \Sigma + \overline \Sigma = e^{-2U},\\
    \delta U: \qquad & \frac{e^{2U}}{6 g^2}\mathcal{D}^\alpha \mathcal{W}_\alpha=\Lambda.
\end{align}
Therefore, we reconstruct \eqref{VeryNice} on-shell, as desired. Plugging these equations of motion back into the Lagrangian, we get precisely \eqref{LG1}, together with the boundary term given in \eqref{BBDDD}.

On the other hand, we can consider the variation of the complex linear superfield $\Sigma$. Actually, since $\Sigma$ is constrained as in \eqref{CLS}, we cannot vary it directly, but we have to take the variation with respect to its prepotential spinor superfield $\Psi_\alpha$, defined as
\be
\Sigma = \overline{\mathcal{D}}_{\dot\alpha} \overline \Psi^{\dot\alpha}, \qquad \overline \Sigma = \mathcal{D}^{\alpha} \Psi_\alpha.
\ee
Notice that, even though $\Sigma$ contains a three-form, when taking its variation we do not have to supplement the Lagrangian with additional boundary terms. Indeed, we have
\be
\begin{aligned}
\delta_{\Psi} {\cal L} & = - \frac34 \int {\rm d}^2 \Theta \, 2 {\cal E} 
 \left( \overline{\cal D}^2 - 8 {\cal R} \right) \, 
 \Lambda \, \overline{\cal D}_{\dot \alpha} \delta \overline\Psi^{\dot \alpha}  + c.c.   
\\
&  =  \frac34 \int {\rm d}^2 \Theta \, 2 {\cal E} 
 \left( \overline{\cal D}^2 - 8 {\cal R} \right) \, 
\overline{\cal D}_{\dot \alpha} \Lambda \, \delta \overline\Psi^{\dot \alpha}  + c.c.  \, , 
\end{aligned}
\ee
which holds without the need to impose non-gauge invariant boundary conditions, since the boundary term produced in the integration by part vanishes identically due to the presence of the chiral projector, namely $\left( \overline{\cal D}^2 - 8 {\cal R} \right) \overline{\cal D}_{\dot \alpha} [ \Lambda \delta \overline\Psi^{\dot \alpha}] \equiv 0$. In other words, the Lagrangian \eqref{Dual1}, as it is, allows for a consistent variation of $\Sigma$. The equations of motion give now
\be
\overline{\mathcal{D}}_{\dot\alpha}\Lambda = 0 = \mathcal{D}_\alpha \Lambda,
\ee
which are solved by
\be
\Lambda = n \, . 
\ee
Inserting this back into \eqref{Dual1}, we get eventually
\be
\label{Freed}
{\cal L} =  \left( \frac{1}{4g^2} \int {\rm d}^2 \Theta \, 2 {\cal E} \, {\cal W}^2(U) + c.c.  \right)  
- 3 n \int {\rm d}^4 \theta \, E \,  e^{-2U} \,. 
\ee
This is the dual formulation of the supergravity theory \eqref{LG1}, where the vector multiplet $U$ is now standard and not given by \eqref{VeryNice}. The Lagrangian \eqref{Freed} is precisely the superspace description of the Freedman model for the Fayet--Iliopoulos D-term in supergravity.  Again, in this formulation, the Hilbert--Einstein term is generated dynamically and is controlled by the flux $n$, which has to be positive for a consistent propagation of gravity.

\subsubsection{Breaking scale invariance with the new Fayet--Iliopoulos term} 
\label{sec:newFI}

As we discussed, the superspace Lagrangian \eqref{LG1} is possibly the simplest one enjoying super-Weyl invariance, before we integrate out the gauge three-form. However, in the minimal model we analysed, we eventually found a degeneracy between the Planck scale and the scale governing the cosmological constant. As we will see in the next section, one way to break such a degeneracy is to couple the model to matter superfields with a non-trivial superpotential. However, in this section, we discuss another possible strategy, which does not require the introduction of new ingredients, but it can be pursued using solely the composite vector multiplet $U$.

We would like now to explicitly break the super-Weyl invariance by introducing an independent supersymmetry breaking scale. Indeed, in the model \eqref{LG1}, supersymmetry and scale invariance were broken at the same stage, by the flux associated to the gauge three-form. To this purpose, we can modify \eqref{LG1} by supplementing it with an additional superspace coupling, which has to be super-Weyl invariant as long as we remain off-shell. Using only the vector superfield $U$, it is possible to introduce the following term
\be
\label{LNEW}
{\cal L}_\text{new FI} = 24 \xi \, \int {\rm d}^4 \theta \, E \, \text{e}^{-2 U} \, \frac{{\cal W}^2 \overline {\cal W}^2 }{{\cal D}^2 {\cal W}^2 \overline {\cal D}^2 \overline {\cal W}^2} \ {\cal D}^\alpha \mathcal{W}_\alpha \, + \, {\cal L}_{bd'} \, , 
\ee
where $\xi$ is a real parameter with $[\xi]$=2 and ${\cal L}_{bd}^\prime$ is an appropriate boundary term, which guarantees the consistent variation of the three-form. We do not give the full superspace form of ${\cal L}_{bd'}$, but we will only consider the parts that are relevant for us. Notice that, in order to consistently introduce such a term, we have to assume that supersymmetry is spontaneously broken by the three-form flux. 

The term \eqref{LNEW} is the super-Weyl invariant version of the new Fayet--Ilioupoulos term in supergravity \cite{Cribiori:2017laj,Antoniadis:2018oeh}. Its bosonic contribution, in the WZ gauge, reads 
\be
\label{LNEWcomp}
{\cal L}_\text{new FI} = - 3 \xi e {\rm D} + {\cal L}_{bd}^\prime = - 3 \xi e \left( \frac{1}{6}R+  *\text{d}C_3-\frac{1}{9}(b_m+3v_m)^2 +\frac19 M\overline M \right) + {\cal L}_{bd}^\prime \, , 
\ee
where the boundary term is
\be
{\cal L}_{bd}^\prime= 3 \xi e  *\text{d}C_3 \, .
\ee
Notice that the three-form effectively drops out from \eqref{LNEWcomp}, when considering also the boundary term. However, to construct a consistent model, we have to add \eqref{LNEW} to the kinetic Lagrangian \eqref{LG1}, which is quadratic in the composite auxiliary field D. Then, when integrating out the three-form from the kinetic part, the Hilbert--Einstein term is generated dynamically, exactly as explained before. Once also the auxiliary fields are integrated out, the contribution of \eqref{LNEWcomp} to \eqref{onshellbulk} is of the type
\be
\label{LNEWonshell}
{\cal L}_\text{new FI} =  - \frac12 \xi e R  \,  
\ee
and it goes together with the Hilbert--Einstein term stemming from \eqref{LG1}. As a result, in the complete theory, the Planck mass and the cosmological constant will depend differently on the two scales $n$ and $\xi$, namely
\be
M_P^2 \sim \xi + \frac{n}{g^2} \, , \quad \Lambda_{cc} \sim \frac{n^2}{g^2} \, . 
\ee
Once we normalize $M_P$ to unity by an appropriate Weyl rescaling of the metric, the cosmological constant is not normalized to $g^2$ and thus does not lose all flux dependence, as it happens in \eqref{onshellbulk-n=1}, instead it becomes
\be
\Lambda_{cc}|_{M_P=1} \sim \frac{n^2}{g^2} \left( \xi + \frac{n}{g^2} \right)^{-2} \sim \left( \frac{g}{n} \xi + \frac{1}{g} \right)^{-2} \, .
\ee 
In other words, by introducing a new scale we made the cosmological constant and the Planck mass independent one from the other.

\subsection{Fayet--Iliopoulos terms and Swampland conjectures}

We showed that the minimal Lagrangian that we proposed for the composite vector multiplet in supergravity is dual to the standard Fayet--Iliopoulos term, described by the Freedman model \cite{Freedman:1976uk}. In this section, which can be also read independently from the rest of the work, we pause our discussion for a moment and we argue that pure Fayet--Iliopoulos terms are in tension with the Weak Gravity Conjecture.\footnote{We would like to thank Thomas Van Riet for discussions related to this subsection.} This result is complementary to earlier no-go theorems on Fayet--Iliopoulos terms in general theories of quantum gravity \cite{Komargodski:2009pc}.

The Lagrangian \eqref{onshellbulk-n=1} for the standard Fayet--Iliopoulos term in supergravity, once the gravitino terms are also included, becomes
\be
\label{freedman}
\begin{aligned}
e^{-1} {\cal L} 
= &   -\frac{M_P^2}{2} R 
+ \frac12 \epsilon^{klmn} \left( \overline \psi_k \overline \sigma_l D_m \psi_n 
- \psi_k \sigma_l D_m \overline \psi_n  \right)  
\\
&  -\frac{1}{4g^2} F_{mn} F^{mn} + i   \epsilon^{klmn} \overline \psi_k \overline \sigma_l \psi_n v_m -  4 g^2 M_P^4 
+ {\cal O}(\lambda) \, , 
\end{aligned}
\ee 
where $D_m$ is the spacetime covariant derivative, which includes the spin-connection $\omega_{ma}^{\ \ \ b}(e,\psi)$, and $ {\cal O}(\lambda)$ refers to terms containing the gaugino, which can be set to zero in the unitary gauge, giving ${\cal O}(\lambda) \equiv 0$. Notice that, with respect to \eqref{onshellbulk-n=1}, we have fixed the gravitino charge to be integer, namely $q_{3/2}=1$. This can be done by rescaling first the vector, $v^m \to \frac{2\sqrt 2}{3}v^m$, and consequently the gauge coupling, $g\to\frac{2\sqrt 2}{3}g$.

Since the scalar potential is a positive constant, we can think of this set up as a toy model for obtaining de Sitter vacua in minimal supergravity, in four dimensions. The model is particularly simple, given the fact that there are no scalars involved. Therefore, we would like to test how robust is such a model against Swampland conjectures \cite{Palti:2019pca} or, alternatively, we wonder whether the pure Fayet--Iliopoulos model can be consistently embedded into quantum gravity. In this respect, we would find evidence for a negative answer, if the Weak Gravity Conjecture \cite{ArkaniHamed:2006dz} is assumed to hold. 
It is important to stress that the model \eqref{freedman} is not ruled out by the no-go theorem in \cite{Komargodski:2009pc} concerning Fayet--Iliopoulos terms in quantum gravity, because in \eqref{freedman} no matter couplings are present, which are instead a central ingredient in such a no-go.\footnote{A second loophole in the argument of \cite{Komargodski:2009pc} appears in the presence of a nowhere vanishing superpotential. In this case, indeed, one can go to a new K\"ahler frame with a gauge invariant K\"ahler potential of the form $G =K+\log|W|^2$ and with unit superpotential. In such a frame, no R-symmetry transformation enters anymore. Another loophole occurs when the Fayet--Iliopoulos parameter is quantized \cite{Seiberg:2010qd,Distler:2010zg}. We thank Eric Sharpe for pointing out the latter possibility to us.}

We start our argument by noticing that the cosmological constant in \eqref{freedman} is  given by 
\be
\label{Lcc}
\Lambda_{ cc} = 2 g M_P \,  
\ee
and it contributes to the vacuum energy as 
\be 
\langle \mathcal{V} \rangle = \Lambda_{ cc}^2 M_P^2 \, . 
\ee
When wondering whether or not this model could be part of the Swampland, an immediate answer would be given by the so-called {\it no de Sitter} conjecture \cite{Danielsson:2018ztv,Obied:2018sgi}, which would promptly exclude it from the Landscape of consistent effective theories. However, the Swampland program is at present an intricate web of statements and conjectures, all related one another. Along with this logic, we can indeed see that the no de Sitter conjecture is not the only one forbidding the existence of \eqref{freedman} in quantum gravity.

The (magnetic) Weak Gravity Conjecture \cite{ArkaniHamed:2006dz,Palti:2019pca} states that, in an effective theory coupled to gravity and with a U$(1)$ gauge symmetry, as it is in the case under consideration, the cut-off is bounded by the gauge coupling $g$, namely
\be
\label{wgc}
\Lambda_{cut-off} \lesssim g M_P. 
\ee
However, in the simple model \eqref{freedman}, the product $g M_P$ is precisely of the order of the cosmological constant $\Lambda_{cc}$ given in \eqref{Lcc}. Therefore, for the standard embedding of the pure Fayet--Iliopoulos term in supergravity, we find that $\Lambda_{cut-off} \lesssim \Lambda_{cc}$, which implies that such an effective description breaks down at the scale given by the vacuum energy, where it is expected to receive non-negligible corrections. From this argument, we see that the model \eqref{freedman} does not give an effective theory consistent with quantum gravity, if the weak gravity conjecture is assumed. On the other hand, such a result is fully compatible with the no de Sitter conjecture.\footnote{See also \cite{Aldazabal:2018nsj} for a discussion on Fayet--Iliopoulos terms from heterotic string and the Swampland, and \cite{Ferrara:2019tmu} for further discussions on the de Sitter conjectures in supergravity.}

Recently, a new embedding of the Fayet--Iliopoulos D-term into supergravity has been proposed \cite{Cribiori:2017laj}. In this setup, the abelian vector is gauging a U$(1)$ symmetry which is not an R-symmetry; therefore, the gravitino is not charged. As a consequence, a new scale $m_{3/2}$ can be inserted into the theory, by means of a constant superpotential. The Lagrangian is 
\be
\begin{aligned}
e^{-1} {\cal L} 
= &   -\frac{M_P^2}{2} R 
+ \frac12 \epsilon^{klmn} \left( \overline \psi_k \overline \sigma_l D_m \psi_n 
- \psi_k \sigma_l D_m \overline \psi_n  \right)  
\\
&  -\frac{1}{4g^2} F_{mn} F^{mn}   -  (4g^2 M_P^4 - 3 m_{3/2}^2 M_P^2) 
+ {\cal O}(\lambda) \,  
\end{aligned}
\ee
and the vacuum energy is given by
\be
\langle \mathcal{V} \rangle = \Lambda_{cc}^2 M_P^2,\qquad \Lambda_{cc} = \sqrt{4 g^2 M_P^2 - 3 m_{3/2}^2}.
\ee
Combing this with the bound \eqref{wgc} imposed by the Weak Gravity Conjecture, up to order one numerical factors we find
\be
\Lambda_{cut-off}\lesssim \sqrt{\Lambda_{cc}^2 + 3 m_{3/2}^2} \sim \Lambda_{SUSY} \, , 
\ee
where $\Lambda_{\text{SUSY}}$ is the supersymmetry breaking scale. This result is clearly different from that obtained in the standard Fayet--Iliopoulos case and, moreover, it is not unexpected. Indeed, due to the underlying non-linear realization of supersymmetry \cite{Cribiori:2017laj}, at the supersymmetry breaking scale the new Fayet--Iliopoulos term is expected to receive corrections. Therefore, in this model, the cut-off is bounded by $\Lambda_{SUSY}$, as the Weak Gravity Conjecture consistently tells us.

Turning the logic around, this argument is in fact showing that, while standard pure Fayet--Ilioupoulos terms are ruled out, the new ones might still be allowed in quantum gravity, if one introduces also a constant contribution in the superpotential. The exact string theory origin of the new Fayet--Iliopoulos terms is therefore an interesting question.

\section{Coupling to supersymmetric membranes}
\label{sec:mem}

In this section, we resume our analysis of the composite three-form gauge vector multiplet in supergravity. We take advantage of the presence of a gauge three-form inside the vector superfield, in order to couple the system to an effective super-membrane in four dimensions \cite{Bergshoeff:1987cm}. First, we present and review how the membrane can be introduced into the setup and then we specify our background of interest and we derive the associated 1/2-BPS flow equations. These will be solved in two different examples in sections \ref{sec:singdw} and \ref{sec:smoothdw}. \footnote{It would be interesting to investigate also backgrounds with strings as those studied for example in \cite{Dvali:2003zh,Lanza:2019xxg,Bandos:2019lps}, but we leave this for future work.}

\subsection{Super-membranes and kappa-symmetry} 

We start by reviewing how to couple a super-membrane to the gauge three-form, following mainly the discussion in \cite{Kuzenko:2017vil,Bandos:2018gjp}. Here, we do not present a complete analysis, rather we focus only on those ingredients which are going to be important for our purposes.

The embedding of the super-membrane world-volume ${\cal C}$, with coordinates $\xi^i$ ($i=0,1,2$), into the four-dimensional $\mathcal{N}=1$ superspace is described by the coordinates
\be
Z^M = ( X^m(\xi^i), \theta^\alpha(\xi^i), \overline \theta_{\dot \alpha}(\xi^i) ) \,. 
\ee 
The pull-back of the super-vielbein $E_M^A$ on the membrane world-volume and the world-volume metric of the membrane are given respectively by 
\be
E_i^A = E_M^A(Z(\xi)) \, \partial_i Z^M(\xi) \ , \qquad h_{ij} = \eta_{ab} E^a_i E^b_j,
\ee
where $\partial_i Z^M = \frac{\partial Z^M}{\partial \xi^i}$ and $\eta_{ab}$ is the four-dimensional flat space metric. Then, a generic kappa-symmetric membrane action has the form 
\be
\label{SMS}
S_{\rm SM} = S_{\rm NG} + S_{\rm WZ} = -2 Q \int_{\cal C} {\rm d}^3 \xi \sqrt{-\det h} \left| T \right| - Q \int_{{\cal C}}{\cal A}_3 \, , 
\ee
where $Q$ is a real positive constant. The chiral superfield $T$, which describes the field-dependent membrane tension, and the real three-form superfield $\mathcal{A}_3$ are both defined in terms of a prepotential superfield $\mathcal{P}$. In particular, the chiral superfield $T$ is given by
\be
\label{TTT}
T = - \frac{i}{4} \left( \overline{\cal D}^2 - 8 {\cal R} \right) {\cal P} \, , 
\ee
while $\mathcal{A}_3$ is defined as
\be\label{super3form}
\begin{aligned}
{\cal A}_{3}=&\,  { -}2 i E^a \wedge E^\alpha \wedge \overline E^{\dot\alpha}  \sigma_{a\;\alpha\dot\alpha}\,{\cal P} + {\frac 12}  E^b\wedge E^a \wedge  E^\alpha
\sigma_{ab\; \alpha}{}^{\beta}{\cal D}_{\beta}{\cal P} \\ &+{\frac 12}  E^b\wedge E^a \wedge  \overline E^{\dot\alpha}
\overline\sigma_{ab}{}^{\dot\beta}{}_{\dot\alpha} \overline{\cal D}_{\dot\beta} {\cal P}   
+\frac{1}{24} 
  E^c \wedge E^b \wedge E^a \epsilon_{abcd} \,\left(\overline{\sigma}{}^{d\dot{\alpha}\alpha}
  [{\cal D}_\alpha, \overline{\cal D}_{\dot\alpha}]{\cal P}  + 8 G^d{\cal P} \right)
 \, . 
\end{aligned}
\ee
If we choose the prepotential to depend on the complex linear superfield $\Sigma$ as
\be
\label{PPP}
{\cal P} = -2 \, {\rm Im} \Sigma \, , 
\ee
such that it transforms under the gauge transformation \eqref{Lgauge} as 
\be 
\label{dcP}
{\cal P} \to {\cal P} + 2 L \, , 
\ee 
then the lowest component of the superfield ${\cal A}_3$ contains the gauge three-form $C_{mnp}$, namely
\be
{\cal A}_{3} \Big{|}_{WZ} = - \frac13 e^c \wedge e^b \wedge e^a \epsilon_{abcd} C^d \, . 
\ee
In addition, the chiral superfield $T$ is invariant under the gauge transformation \eqref{dcP} and transforms under super-Weyl as 
\be
T \to e^{-6 Y} T \, . 
\ee 
Keeping all of these transformations into account, one can check that the Nambu--Goto and Wess--Zumino terms in \eqref{SMS} are super-Weyl and gauge invariant. This is a crucial requirement for our construction, since it is needed for the consistency of the R-symmetry gauging.

As it is known, the full action \eqref{SMS} has a fermionic symmetry called kappa-symmetry, under which the coordinates transform as 
\be
\delta Z^M(\xi) = \kappa^\alpha E_\alpha^M + \overline \kappa_{\dot \alpha} \overline E^{\dot \alpha M}  \, , 
\ee
where the local fermionic parameter satisfies the projection \footnote{We would like to thank Dmitri Sorokin for discussions related to kappa-symmetry.}   
\be
\label{kappa}
\kappa_\alpha 
=  \frac{T}{|T|} 
\frac{i \tilde \epsilon^{ijk}}{3! \sqrt{-\det h}} \epsilon_{abcd} E^b_i E^c_j E^d_k \sigma^a_{\alpha \dot \alpha} 
\overline \kappa^{\dot \alpha} \, . 
\ee
As a result, only half of the kappa-symmetry parameters are independent, which means that only half of the fermionic world-volume degrees of freedom $\theta^\alpha(\xi)$ are independent as well. A complete and detailed proof of the kappa-symmetry invariance of the action can be found for example in \cite{Bandos:2018gjp}.

For completeness, we write the bosonic sector of the action \eqref{SMS}. To this purpose, we notice that the bosonic components of the chiral superfield $T$ defined in \eqref{TTT}, in the WZ gauge, are given by (only bosons)
\be
\label{Tcomp}
\begin{aligned}
T \Big{|}_{WZ} &=  \overline F\Big{|}_{WZ} - \frac16 M = - \frac13 M \, ,\\
-4F^T\Big{|}_{WZ} &\equiv\mathcal{D}^2 T \Big{|}_{WZ} = -\frac43 i e^m_a \mathcal{D}_m b^a + 4 i \mathcal{D}_m \overline P^m\Big{|}_{WZ} + \frac 43 M \overline M.
\end{aligned}
\ee 
The component form of the super-membrane action \eqref{SMS} is then
\be
\label{SMScomp}
S_{\rm SM} = -\frac23 Q  \int_{\cal C} {\rm d}^3 \xi |M| \sqrt{-\det h} 
- \frac13 Q \int_{{\cal C}} {\rm d}^3 \xi 
\tilde \epsilon^{ijk} \frac{\p X^m}{\p \xi^i} \frac{\p X^n}{\p \xi^j} \frac{\p X^p}{\p \xi^k} C_{mnp}   
\, + \text{fermions} 
\, ,
\ee
where we remind that $M$ is the supergravity auxiliary field and $h_{ij}\equiv g_{mn}\partial_i X^m \partial_j X^n$ is now the pull-back metric on the membrane. Our conventions for the values of the antisymmetric symbols are $\tilde \epsilon_{012}=-\tilde\epsilon^{012} = -1 = \tilde \epsilon_{0123}=-\tilde\epsilon^{0123}$, as in \cite{Wess:1992cp}. The Levi-Civita tensors are defined as usual, namely 
\be
\epsilon^{ijk} = \frac{\tilde \epsilon^{ijk}}{\sqrt{- \det h_{ij}}} 
\ , \qquad 
\epsilon^{klmn} = \frac{\tilde \epsilon^{klmn}}{e} \, , 
\ee
and we also recall the relations
\be
\label{Cform}
C^m = \frac{1}{3!} \epsilon^{mnpq} C_{npq} ,\qquad C_{mnp} = \epsilon_{mnpq} C^q ,\qquad \partial_{[m} C_{npq]} = -\frac14 \epsilon_{mnpq}*{\rm d}C_3,
\ee
which are being employed through the work. For example, by using the third one, we can recast the equation of motion of $X^m$ stemming from \eqref{SMScomp} as
\be
\label{eomXm}
\partial_m \left(|T| \sqrt{-\det h}\right) - \frac16 \tilde \epsilon^{ijk} \partial_i X^n \partial_j X^p \partial_k X^q \epsilon_{mnpq} *{\rm d}C_3=0.
\ee

Finally, without loss of generality, we can choose the membrane to sit at $z=0$ and we can align its world-volume coordinates with the remaining spacetime ones, namely
\be
\label{staticgauge}
{\text{transverse:}} \quad z \ , \qquad {\text{world-volume:}} \quad \xi^i \equiv x^i = (t,x,y) \, . 
\ee
This is the so-called static gauge.

\subsection{Metric ansatz and BPS flow equations}
\label{subsec:BPSeq}
We present now a simple metric ansatz which will be used as a background in models with membranes. Then, we derive the associated first order 1/2-BPS flow equations, which are obtained by setting the supersymmetry variations of the fermions to zero, on such a background.

We are interested in studying domain walls separated by supersymmetric membranes in a background of the type
\be
\label{metric4d}
ds^2 = e^{2A(z)}(-{\rm d}t^2+{\rm d}x^2+{\rm d}y^2) + {\rm d}z^2,
\ee
where $A(z)$ is a warp factor which depends only on the coordinate $z$. 
This metric is compatible with the presence of a flat rigid membrane at the $z=0$ position. 
In the basis of vielbein one-forms $\{e^a\}$, given by
\be
e^0 =  {\rm d}t\,\, e^A, \qquad e^1 =  {\rm d}x\,\, e^A, \qquad e^2 =  {\rm d}y\,\, e^A, \qquad e^3 ={\rm d}z,
\ee
the only non vanishing components of the spin connection one-form, $\omega^{ab}$, are
\be
\omega^{i3} = -e^i \,\,\dot A , \qquad \text{i.e.}\qquad  {\omega_m}^{i3} = -e^A \dot A\, \delta_m^i, \qquad i=0,1,2,
\ee
where a dot means derivative with respect to $z$, namely $\dot A \equiv \partial_z A$.
In turn, the only non vanishing components of the curvature two-form, $R^{ab}={\rm d}\omega^{ab}+\omega^{ac}{\omega_c}^b$, are given by
\be
R^{i3} = - e^i e^3(\ddot A +\dot A^2), \qquad R^{ij} = - e^i e^j \dot A^2,
\ee
from which the Ricci scalar, $R =e^m_ae^n_b{R_{mn}}^{ab}$, is found to be
\be
\label{Ricci}
R = 6 \ddot A + 12 \dot A^2.
\ee

Since we are interested in a purely bosonic background, in order to preserve supersymmetry we have to set to zero the supersymmetry variations of the fermions. In addition to the spin-3/2 gravitino, in $\mathcal{N}=1$ supergravity we have two classes of spin-1/2 fermions, namely those in vector multiplets, $\lambda_\alpha$, and those in chiral multiplets, $\chi_\alpha^I$.\footnote{We consider just one vector multiplet, but a generic number of chiral multiplets $\Phi^I$.} Therefore, we consider the following equations
\be
\label{flow}
\begin{aligned}
\delta \lambda_\alpha: & \quad  0 = e_a^m e_b^n F_{mn} \sigma^{ab}{}_\alpha{}^\rho \xi_\rho + i \xi_\alpha {\rm D} \, , 
\\
\delta \chi_\alpha^I: & \quad 0 = i \sigma^m_{\alpha \dot \beta} \overline \xi^{\dot \beta} \tilde{\cal D}_m \Phi^I + F^I \xi_\alpha \, , 
\\
\delta \psi_{m \alpha}: & \quad 0 =- 2 \tilde D_m \xi_\alpha + \frac{i}{3} M \, e_m^a  \sigma_{a \alpha \dot \alpha} \overline \xi^{\dot \alpha} + \text{terms with}\, b_a \, , 
\end{aligned}
\ee
where, in our setup, D is the composite auxiliary field given in \eqref{Dcomp}. The gauge covariant derivative acting on the scalars is defined as
\be
\tilde{\cal D}_m \Phi^I = \partial_m \Phi^I - \frac 12 X^I \, v_m \, , 
\ee
where $X^I$ is the holomorphic Killing vector, which is defined by matching with the gauge transformation of the scalar $A^I$ under the gauged $U(1)$, 
that is: $\delta A^I = \epsilon X^I$. 
Once we study a specific K\"ahler manifold then we can derive the Killing potential $\mathbb{D}$ that is real and is given by 
\be
X^I = - i g^{I \overline J} \partial_{\overline J}\mathbb{D} \, , 
\ee
where $g_{I \overline J}$ is the K\"ahler metric. 
For the supersymmetry parameter we have 
\be
\tilde D_m \xi^\alpha = \partial_m \xi^\alpha + \xi^\rho \omega_{m \rho}{}^{\alpha}, \qquad \text{with} \qquad 
\omega_{m \rho}{}^{\alpha} = -\frac12 \sigma^{ab}{}_{\rho}{}^{\alpha} \omega_{mab} \, . 
\ee
In order for the solution of \eqref{flow} to preserve part of the bulk supersymmetry (BPS solution), we have to impose a restriction on the supersymmetry parameter $\xi_\alpha$, such that not all of its components are independent. The appropriate condition is of the form
\be
\label{xiproj}
\xi_\alpha = e^{i\beta} \sigma^3_{\alpha \dot \alpha }\overline \xi^{\dot\alpha},
\ee
where $e^{i\beta}$ is a constant phase. Indeed, as a consequence of \eqref{xiproj}, only two out of four supercharges are independent. This condition is similar in form to the kappa-symmetry projection \eqref{kappa}. Indeed, at $z=0$ we will align the fermionic kappa-symmetry parameter $\kappa_\alpha$ with the supersymmetry parameter $\xi_\alpha$, in order that the membrane preserves part of the bulk supersymmetry.

It is important to notice that, since we have a non-vanishing D-term generated by the U$(1)_R$ gauging, the first equation in \eqref{flow} would need a non-trivial profile for the vector $v_m$, in order to be solved (we exclude the trivial case in which $\xi_\alpha=0$). Therefore, to avoid possible complications related to anisotropies induced by a non-vanishing vacuum-expectation value of $v_m$, we require additionally that
\be
\label{D=0}
{\rm D} \equiv \frac{1}{6}R+  *{\rm d}C_3 -\frac{1}{9}(b_m+3v_m)^2  +\frac19 M \overline M = 0 
\ee
along the flow, in order that we can consistently set 
\be
\label{novect}
v_m =0 = b_m,
\ee
where the second equality follows from gauge invariance. With these conditions, the first equation in \eqref{flow} is automatically solved, without posing any restriction on the supersymmetry parameter.

Inserting \eqref{metric4d}, \eqref{xiproj}, \eqref{D=0} and \eqref{novect} into \eqref{flow} and assuming that all the quantities depend only on $z$, in particular $\Phi = \Phi(z)$ and $\xi_\alpha = \xi_\alpha(z)$, we obtain the following set of BPS flow equations
\begin{align}
\label{BPS1}
    \dot \Phi^I &= i e^{i\beta} F^I,\\
\label{BPS2}
    \dot A &= \frac i3 e^{-i\beta} M,\\
\label{BPS3}
    \dot \xi_\alpha &=\frac{\dot A}{2}\xi_\alpha.
\end{align}
The last equation can be directly solved to give 
\be
\xi_\alpha(z) = e^{\frac{A(z)}{2}}\tilde \xi_\alpha \ , \qquad \tilde \xi_\alpha = {\rm constant} \, ,
\ee
while \eqref{BPS1} and \eqref{BPS2} will be solved in the next sections, within two different examples.

\section{Irregular BPS domain walls} 
\label{sec:singdw}

In this section, we present the first of our examples, in which the composite three-form vector superfield is coupled to one charged matter chiral superfield. We start by finding supersymmetric AdS vacua and then we introduce a supersymmetric membrane into the setup, following the lines of section \ref{sec:mem}. When looking for BPS domain wall solutions interpolating between such vacua, we notice that there are various irregularities, which might signal the fact that the solution itself is not physical. We have however decided to keep this example in our presentation in order to show what difficulties may arise. In section \ref{sec:smoothdw}, we will later present a possible strategy to cure this problem, within a second example.

\subsection{Bulk Lagrangian with one charged chiral superfield} 
We start by presenting the matter coupled model without the membrane and by finding supersymmetric AdS vacua. This model will capture the physics in the bulk region, away from the membrane source, once the latter will be introduced.

Given the general form of the scalar potential in gauged supergravity,
\be
\mathcal{V} = e^{K / M_P^2} \left(|DW|^2 - 3 M_P^{-2} |W|^2\right) + \frac18 g^2 \mathbb{D}^2,
\ee
where $\mathbb{D}$ are the killing potentials associated to the gauged isometries of the scalar manifold, in order to have AdS vacua the superpotential $W$ has to get a non-vanishing vacuum-expectation-value. A simple choice would then be a constant superpotential, but this would break explicitly the U$(1)$ R-symmetry that we are currently gauging. To avoid the issue, we can take $W$ to be a function of a chiral superfield $S$
\be
S = S + \sqrt 2 \Theta \chi^S + \Theta^2 F^S \, , 
\ee
transforming under super-Weyl (namely gauge) transformations  as 
\be
S \to e^{-qY} S. 
\ee
In this way, indeed, a non-trivial superpotential can be included in the theory and the gauged R-symmetry is only spontaneously broken in the vacuum.

A simple, super-Weyl invariant action describing the couplings of the composite vector superfield $U$ to the charged chiral superfield $S$ is
\be
\label{model-q}
\begin{aligned}
{\cal L} = & - 3 \int {\rm d}^4 \theta E \, S\,e^{(q-2) U}  \overline{S} 
 + \left( \int {\rm d}^2 \Theta \, 2 {\cal E} \, f \, S^{\frac 6q} + c.c.  \right) 
\\
& 
+ \left( \frac{1}{4g^2} \int {\rm d}^2 \Theta \, 2 {\cal E} \, {\cal W}^2(U) + c.c.  \right)  
+ {\cal L}_{bd} \, , 
\end{aligned}
\ee
where $f$ is a real constant parameter, with mass dimensions 
\be 
[f]=3 - \frac 6q \, , 
\ee 
and is at this point the only dimensionful parameter in the action. 
Indeed, as we explained in the previous section, $M_P$ will appear only once the 3-form is integrated out and the associated flux enters. Notice that, since we have now introduced new couplings to $U$, which is not a gauge-invariant quantity, the boundary term is expected to be different from \eqref{BBDDD}. Nevertheless, one can calculate it along with the same logic as before. Starting from the following parent Lagrangian 
\be
\begin{aligned}
\label{dualmodelq}
{\cal L} = & -3 \int {\rm d}^4 \theta E \, S\,e^{(q-2) U}  \overline{S} 
 + \left( \int {\rm d}^2 \Theta \, 2 {\cal E} \, f \, S^{\frac 6q} +\frac{1}{4g^2} \int {\rm d}^2 \Theta \, 2 {\cal E} \, {\cal W}^2(U) + c.c.  \right)\\
 &- \frac34 \left( \int {\rm d}^2 \Theta \, 2 {\cal E} 
 \left( \overline{\cal D}^2 - 8 {\cal R} \right) \, 
 \Lambda \left( \Sigma - \frac12 e^{-2U} \right) + c.c. \right) \,,
 \end{aligned}
\ee
one can take the variation with respect to the Lagrange multiplier real superfield $\Lambda$, which gives again \eqref{VeryNice}, and with respect to the unconstrained $U$, resulting in
\be
\delta U:\qquad \frac{e^{2U}}{6 g^2}\mathcal{D}^\alpha \mathcal{W}_\alpha  + \frac{(q-2)}{2}e^{qU}S\overline S = \Lambda.
\ee
Plugging these back into the Lagrangian \eqref{dualmodelq}, we obtain \eqref{model-q} with the boundary term
\be
\label{Lbdex1}
{\cal L}_{bd} = - \frac{1}{16 g^2}  \int {\rm d}^2 \Theta \, 2 {\cal E} \left( \overline{\cal D}^2 - 8 {\cal R} \right) 
\Big{[} \left(q-2 \right) 
\frac{3 g^2 S \overline S}{( \Sigma + \overline \Sigma )^{q/2} }
+   \, 
\frac{{\cal D}^\alpha {\cal W}_\alpha }{\Sigma + \overline \Sigma} \Big{]} ( \Sigma - \overline \Sigma ) + c.c. 
\ee
Its bosonic component expansion, in the WZ gauge, is
\be
e^{-1}{\cal L}_{ bd}  = \frac{1}{g^2} e_a^n D_n \left[ \left( -\frac{R}{6} 
- *\text{d}C_3 
+ \frac{1}{9}(b_m+3v_m)^2  
- \frac{ M \overline M }{9} 
+  \frac{3g^2 (q-2)}{2}  S \overline S \right) C^a \right]  \, . 
\ee

In order to study the properties of the model, one can either start from the theory \eqref{model-q} containing the three-form, or equivalently use the dual formulation, which can be constructed by integrating out $\Sigma$ from \eqref{dualmodelq}. For convenience, we decide to follow this second path. 
The integration of $\Sigma$ proceeds then as before and gives again a constant, $\Lambda=-m$. Once this is inserted back into \eqref{dualmodelq}, the Lagrangian becomes 
\be
\label{model-q2}
\begin{aligned}
{\cal L} = & -3 \int {\rm d}^4 \theta E \, S\,e^{(q-2) U} \,\overline S 
 + \left( \int {\rm d}^2 \Theta \, 2 {\cal E} \, f \, S^{\frac 6q} + c.c.  \right)
\\
& 
+ 3 m \int {\rm d}^4 \theta \, E \,  e^{-2U} 
+ \left( \frac{1}{4g^2} \int {\rm d}^2 \Theta \, 2 {\cal E} \, {\cal W}^2(U) + c.c.  \right)   \, . 
\end{aligned}
\ee
This is a standard superspace Lagrangian for gauged supergravity \cite{Wess:1992cp}, with\footnote{We set $M_P=1$ from now on except otherwise noted.} 
\be
\label{KGamma}
K + \Gamma = -3 \log \left(  S e^{(q-2) U}  \overline S - m e^{-2U} \right) \ , \quad W = f \, S^{\frac{6}{q}} \,.  
\ee
Here, we stress that $U$ is now a standard vector superfield and not the composite object \eqref{VeryNice}. From \eqref{KGamma}, we find that the K\"ahler potential and the Killing potential are given respectively by
\be
K = -3 \log \left( S \overline S-m  \right) \ , \qquad 
\mathbb{ D} = \frac{6 m + 3 (q-2) \, S\overline S}{m-S\overline S}  \, .
\ee
Notice that positiveness of the K\"ahler metric 
\be
g_{S\overline S}  = \frac{3m}{(S\overline S-m)^2} \, , 
\ee
requires 
\be
m>0 \, , 
\ee 
while the moduli space is bounded to be within 
\be
\label{modspace}
S\overline S > m.
\ee
After the standard Weyl rescaling \cite{Wess:1992cp}, the component bosonic sector becomes
\be
e^{-1}\mathcal{L} = -\frac{1}{2} R - g_{S \overline S} {\tilde {\cal D}}_m S {\tilde {\cal D}}^m \overline S-\frac{1}{4 g^2}F_{mn}F^{mn} - V(S,\overline S),
\ee
where the gauge covariant derivative is 
\be
{\tilde {\cal D}}_m S = \partial_m S +\frac i2 g^{S \overline S}\mathbb{D}_{\overline S}v_m.
\ee 
The scalar potential reads
\be
\label{Vqmodel}
\mathcal{V}(S,\overline S) = 
\frac{1}{( S\overline S -m)^3} \left( \frac{(S\overline S-m)^2}{3 m} |D_S W|^2 - 3 f^2 (S \overline S)^{6/q} \right)+\frac{ g^2}{8} \mathbb{ D}^2  \,,
\ee
where
\be
D_S W = 3 f S^{\frac 6q -1}\left(\frac 2q - \frac{S\overline S}{S\overline S-m}\right) \, . 
\ee

We now study the existence of supersymmetric AdS vacua in the model under investigation. To find them, we can solve the F-term and D-term equations
\be
\label{susypoint}
D_S W  = 0,\qquad \mathbb{D}=0.
\ee
Changing coordinates, $S = \rho e^{{i} \theta}$, the solution of both equations is
\be
\label{AdSvacua}
S \overline S = \rho^2 = \frac{2m}{2 - q}, \qquad \text{with} \qquad q<2.
\ee
Notice that the fact that F-term and D-term equations are solved simultaneously is expected, since it is known that, in supergravity, F-term and D-term breaking are related. Furthermore, a solution of \eqref{susypoint} is always an extremum of the potential, namely $\partial_S V =0$. The vacuum energy at \eqref{AdSvacua} is then given by
\be
\langle \mathcal{V} \rangle=-\frac{3 f^2}{m^3} \left(\frac{2-q}{q}\right)^3\left(\frac{2m}{2-q}\right)^{\frac 6q} \equiv - 3 m_{3/2}^2\, , 
\ee
and it is negative for $2 > q >0$, namely for these values such vacua are AdS. Furthermore, we defined the Lagrangian gravitino mass as $m_{3/2} = e^{\frac K2}|W|$. 

The scalar field $\theta$ is massless, while $\rho$ is massive and its canonically normalized mass is given by
\be
m^2_\rho = 3g^2(2-q)-\frac{2 f^2}{m^3} \left(\frac{2-q}{q}\right)^3\left(\frac{2m}{2-q}\right)^{\frac 6q}\equiv 3 g^2(2-q)-2 m_{3/2}^2.
\ee
Due to the minus sign between the first and the second term, this mass can assume positive and negative values as well. However, we recall that a negative Lagrangian mass for a scalar field in AdS does not lead to instabilties, as long as it satisfies the Breitenlohner--Freedman bound \cite{Freedman:2012zz}
\be
m^2_\rho > \frac34 \mathcal{V} \, . 
\ee
Indeed, this is the case for the model under investigation, since
\be
m^2_\rho - \frac34 \langle \mathcal{V} \rangle = 3 g^2(2-q)+\frac{m_{3/2}^2}{4} >0.
\ee
Since the gauged U$(1)$ R-symmetry is spontaneously broken on this vacuum, due to a non-vanishing $f$, the abelian vector acquires a mass by means of the Brout--Englert--Higgs mechanism 
\be
m_v^2 = \frac{g^2}{2} g^{S\overline S}\mathbb{D}_S\mathbb{D}_{\overline S} \Big{|}_{S \overline S =\rho^2}= 3g^2 (2 -q) \, .
\ee

Finally, notice that the AdS vacua we find here are in agreement with the magnetic Weak Gravity Conjecture, since the vacuum energy is independent of the gauge coupling $g$ and therefore can be chosen to satisfy \eqref{wgc}. Moreover, we also see that $\langle W \rangle \ne 0$, therefore we might go to an appropriate K\"ahler frame such that the no-go of \cite{Komargodski:2009pc} is evaded. Indeed, these vacua are supersymmetric AdS and therefore are not expected a priori to be in the Swampland.

\subsection{Membrane coupling and BPS flow equations}

We would like now to couple the system to a supersymmetric membrane and derive the BPS flow equations associated to background presented in section \ref{sec:mem}. To this purpose, we add to \eqref{model-q} the membrane action \eqref{SMS}, obtaining
\be
\label{compex1}
\begin{aligned}
S =   \int d^4 x \Bigg{(} & \ -3 \int {\rm d}^4 \theta E \, S\,e^{(q-2) U}  \overline{S}  + \left( \int {\rm d}^2 \Theta \, 2 {\cal E} \, f \, S^{\frac 6q} + c.c.  \right) \\ 
& + \left( \frac{1}{4g^2} \int {\rm d}^2 \Theta \, 2 {\cal E} \, {\cal W}^2(U) + c.c.  \right)  + {\cal L}_{bd} \Bigg{)}  \\ 
-2 Q  \int_{\cal C} &  {\rm d}^3  \xi \sqrt{-\det h} \left| T \right| - Q \int_{{\cal C}}{\cal A}_3 \, , 
\end{aligned}
\ee
where the boundary term is given in \eqref{Lbdex1}. For convenience, we change again basis to $S=\rho e^{i\theta}$ and, as explained in \ref{subsec:BPSeq}, in the following we systematically set $v_m=b_m=0$, since we are interested in isotropic domain wall solutions. Indeed, we will also check that \eqref{D=0} is always satisfied along the flow. With these assumptions, the bosonic component form Lagrangian reduces to
\be
\begin{aligned}
e^{-1}\mathcal{L}&=-\frac12 \rho^2 R + 3 \partial_m (\rho e^{i\theta})\partial^m (\rho e^{-i\theta})\\
&-\frac13 \rho^2 M\overline M + \rho e^{i\theta}\overline M \overline F^S+\rho e^{-i\theta}M F^S - 3 F^S \overline F^S\\
&-f(\rho e^{i\theta})^{\frac 6q}\overline M-f(\rho e^{-i\theta})^{\frac 6q} M+\frac 6q f (\rho e^{i\theta})^{\frac 6q -1}F^S+\frac 6q f (\rho e^{-i\theta})^{\frac{6}{q} -1}\overline F^S\\
&-\frac32 (q-2)\rho^2{\rm D}+\frac{1}{2g^2} {\rm D}^2\\
&+ e^{-1}\int {\rm d}^3\xi \delta^{(4)}(x-X)\left(-\frac23 Q |M|\sqrt{-\det h}-\frac Q3 \tilde \epsilon^{ijk}\partial_i X^m \partial_j X^n \partial_k X^p C_{mnp}\right).
\end{aligned}
\ee 
Here D is given by \eqref{Dcomp} with $v_m=0=b_m$. 
The variation of the auxiliary fields, assuming \eqref{D=0}, give
\be
\begin{aligned}
\label{EOMaux1}
F^S &= \frac{4}{q(q-2)}f \left(\rho e^{-i\theta}\right)^{\frac 6q -1}(e^{i\theta}-1)\\
&-e^{-1} \frac{2Q}{3(q-2)}\rho^{-1}e^{i\theta}\int {\rm d}^3\xi \delta^{(4)}(x-X)  \frac{\overline M}{|M|}\sqrt{-\det h},\\
M &= -\frac{6}{q-2}f\rho^{-2}\left(\rho e^{i\theta}\right)^\frac{6}{q}\left(1-\frac 2q e^{-i\theta}\right)\\
&- e^{-1} \frac{2Q}{q-2}\rho^{-2}\int {\rm d}^3\xi \delta^{(4)}(x-X) \frac{M}{|M|}\sqrt{-\det h}.
\end{aligned}
\ee
It is important to notice that, since $T|_{WZ} = -\frac13 M$, the supergravity auxiliary field enters the membrane tension and modifies non-trivially the equations of motion. Furthermore, in this example we see that both equations \eqref{EOMaux1} have a source term with a delta function on the right hand side. This fact will have important consequences on the nature of the domain walls.

In order to simplify the setup, we assume that all fields depend only on the coordinate $z$. In the static gauge, to which we restrict from now on, the equation of motion of the three-form is then
\be
\label{m}
\frac32 (2-q)\rho^2 = 3m,\qquad \text{with}\qquad m=c_0 + \frac23 Q \Theta(z),
\ee
where $c_0$ is an integration constant, while the variation of $X^m$ gives
\be
*{\rm d}C_3 = \frac13 e^{-3A}\partial_z (e^{3A}|M|).
\ee
We can simplify further the analysis by assuming that
\be 
e^{i\theta}\equiv 1,\qquad F^S=\overline F^S \equiv \mathcal{F},\qquad M = \overline M \equiv \mathcal{M}, 
\ee 
and by taking the charge of the chiral superfield to be unity,
\be 
q=1 \, . 
\ee 
As a consequence of all of these assumptions, the equations of motion of the auxiliary fields reduce to
\begin{align}
\label{EOMF1}
\mathcal{F} &= \frac23 \rho^{-1}Q\, \text{sgn}(\mathcal{M})\delta(z),\\
\label{EOMM1}
\mathcal{M} &= - 6 f \rho^4+2 \rho^{-2}Q \,\text{sgn}(\mathcal{M})\delta(z),
\end{align}
while the variation of $C_3$ and $X^m$ are given respectively by
\begin{align}
\label{solrho}
\rho^2 &=  2 c_0 + \frac43 Q\, \Theta(z),\\
\label{solC3}
*{\rm d}C_3 &= \frac13 e^{-3A}\partial_z (e^{3A} |\mathcal{M}|),
\end{align}
where $\Theta(z)$ is the Heaviside step function.

Since the membrane is supersymmetric, it can preserve part of the bulk supersymmetry on its world volume.\footnote{We recall that a minimal spinor in four dimensions has four real supercharges, while in three dimensions it has only two supercharges. Therefore, on the three-dimensional membrane world volume we can preserve two of the four supercharges of the bulk.} In order for this to happen, we have to require that the bulk supersymmetry parameter $\xi_\alpha(z)$ is aligned with the kappa-symmetry parameter $\kappa_\alpha$ at $z=0$. Since the latter satisfies \eqref{kappa}, on $\xi_\alpha(z)$ we impose that
\be
\xi_\alpha = \frac{T}{|T|} \left( \frac{i \tilde \epsilon^{ijk}}{3! \sqrt{-\det h}} \epsilon_{abcd} E^b_i E^c_j E^d_k \sigma^a_{\alpha \dot \alpha} \right) \Big{|}_{z=0} \overline \xi^{\dot \alpha} =-\frac{i M}{|M|} \sigma^3_{\alpha \dot \alpha} \overline \xi^{\dot \alpha} = -i\,\text{sgn}(\mathcal{M}) \sigma^3_{ \alpha \dot \alpha} \overline \xi^{\dot \alpha}  \, . 
\ee
By comparing this with the projection condition on $\xi$ given in \eqref{xiproj}, we get
\be
e^{i\beta}\equiv -i\frac{M}{|M|}=-i\, \text{sgn}(\mathcal{M}).
\ee
For simplicity, in the following we will also assume that 
\be
\text{sgn}(\mathcal{M})=1,\qquad e^{i\beta}=-i \, . 
\ee
Taking all this into account, the BPS flow equations \eqref{BPS1} and \eqref{BPS2} in the background \eqref{metric4d} reduce to
\begin{align}
\label{bps1}
    \dot \rho &= \mathcal{F},\\
\label{bps2}
    \dot A & =-\frac{\mathcal{M}}3,
\end{align}
where $\mathcal{F}$ and $\mathcal{M}$ are given respectively in \eqref{EOMF1} and \eqref{EOMM1}. As a consistency check, one can verify that the condition \eqref{D=0} is satisfied along the flow. Indeed, using \eqref{Dcomp}, \eqref{Ricci}, \eqref{novect} and \eqref{solC3}, we have
\be
\begin{aligned}
{\rm D} &\equiv \frac16 R + *{\rm d}C_3 + \frac19 \mathcal{M}^2 = \left(\partial_z + 2 \dot A + \frac{\mathcal{M}}{3}\right)\left(\dot A + \frac{\mathcal{M}}3\right)=0,
\end{aligned}
\ee
which vanishes when using also \eqref{bps2}.

It remains to solve the equations \eqref{bps1} and \eqref{bps2}. From their very form, we can see that if the auxiliary fields $\mathcal{F}$ and $\mathcal{M}$ are discontinuous at one point, say $z=0$, then the derivatives $\dot \rho$ and $\dot A$ of the propagating fields are discontinuous but in principle the fields themselves are not. This is for example what happens in \cite{Ceresole:2006iq,Bandos:2018gjp}.  However, in our case, the auxiliary fields contain Dirac delta functions at the position of the membrane. This implies that the fields $\rho$ and $A$ will be discontinuous, as we will see in the explicit solution.

\subsection{Domain wall profile}
\begin{figure}[ht]
\centering 
  \includegraphics[scale=1]{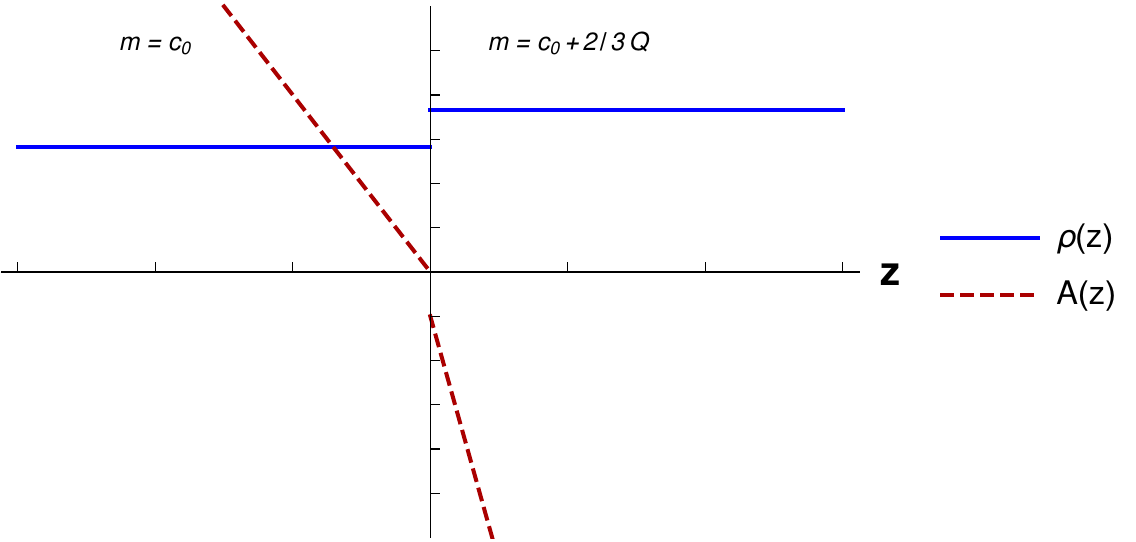}
\caption{ 
{\it Plot of the solutions \eqref{flowa} and \eqref{rpm} for the scalar $\rho(z)$ and the warp factor $A(z)$, with parameters $c_0  =1$, $Q=2$ and $f=-\frac12$. We see that both fields are discontinuous at the position of the membrane.} \label{plotz}} 
\end{figure}

We would like now to solve the BPS flow equations \eqref{bps1} and \eqref{bps2}, namely
\begin{align}\
\label{BPSex1}
    \dot \rho &= \frac23 \rho^{-1}Q \delta(z),\\
    \label{BPSex2}
    \dot A & =2f\rho^4 - \frac23 \rho^{-2}Q \delta(z).
\end{align}
Actually, one can easily check that the first of these equations is identically solved by \eqref{solrho}. We see then that the field $\rho$ experiences a discontinuity when crossing the membrane, as anticipated before. We look now at the equation \eqref{BPSex2} and we determine the profile of $A$. In order to find the discontinuity $\Delta A$ due to the presence of a delta function in \eqref{BPSex2}, we integrate this equation on a small interval $[-\epsilon,\epsilon]$ and subsequently we send $\epsilon$ to zero. We have
\be
\label{discA}
\Delta A = \lim_{\epsilon\to0} \int_{-\epsilon}^{+\epsilon} \left( 2 f \rho^{4} - \frac23 \rho^{-2} Q \delta(z)  \right) dz  = - \frac23 \rho^{-2}(0) Q \, , 
\ee
where $\rho^2(0) = 2 c_o +  \frac43 Q \Theta(0)$, with $\Theta(0)=\frac12$. We solve then the flow equation on the right and on the left sides of the membrane and we obtain
\be
\label{flowa}
A(z) =  \left\{\begin{array}{cc}
     2 z f \rho^{4}_-  & z<0,\\[.3cm] 
     2 z f \rho^{4}_+  - \frac23 \rho^{-2}(0) Q & z>0 ,\\
\end{array}\right.  
\ee
where $\rho_{\pm}(z)$ are given by 
\be
\label{rpm}
\rho(z) :  \left\{\begin{array}{cc}
     \rho_-(z) = \sqrt{2 c_o} & z<0, \\[.3cm]
     \rho_+(z) =  \sqrt{2 c_o +  \frac43 Q}, & z>0  . 
\end{array}  \right.
\ee
A plot of this solution is given in figure \ref{plotz}. In particular, the membrane sits at $z=0$ and discontinuities in the profile of the fields are clearly visible. Away from the membrane, the flow is essentially trivial.

The profile of the domain wall has various irregularities, which we now comment on. First, when going on-shell with respect to the three-form, the Hilbert--Einstein term is normalized as
\be
{\cal L}_{\text{HE, on-shell}} = - \frac12 m e R \, , 
\ee
where $m$ is given in \eqref{m} and it has to be positive. Therefore, we see that the effective $M_P^2$ we discussed in \eqref{MP2} can be identified as $n=m$.  However, in the present solution $m$ is discontinuous. As a result we are not sure on how to study the theory in the Einstein frame and correctly evaluate the curvature. Second, the metric profile in the Jordan frame is characterized by $A(z)$ which is again discontinuous. Such discontinuity could signal the break down of supergravity on the membrane. Finally, both the Einstein equations and the equations of motion for the scalar are not satisfied at $z=0$, if we use the $A$ and $\rho$ profile we have described here. For example, the equation for $\rho$ reduces to 
\be
\rho \mathcal{M} (\mathcal{M} + 6f\rho^4)=0,
\ee
which is formally not satisfied by \eqref{EOMM1} at $z=0$. For these reasons we do not investigate this profile 
further.\footnote{An interpretation of these irregularities is that the current system has no supersymmetric ground state at all (even BPS), since the equations of motion are formally not solved on the membrane. However, since we did not investigate whether or not the system allows for any other type of BPS solution, we remain agnostic in this respect.} Let us however point out that the root of the above issues is the fact that we have a Dirac delta function appearing in the auxiliary field $M$. We will now see how to overcome this issue.

\section{Smooth BPS domain walls} 
\label{sec:smoothdw}

In the previous section we studied BPS domain walls where the physical fields are discontinuous when crossing the membrane. Moreover, at the membrane position, various irregularities are present which might signal the breakdown of supergravity. The very origin of this problem is rooted in the fact that the membrane couples to an auxiliary field, which changes non trivially the BPS equations, even if is not propagating. In this section, we let the auxiliary field propagate. 
This will introduce higher derivatives, albeit ghost-free, into the theory and, as we will see, it will also lead to smooth domain wall solutions. 

For convenience of the presentation, we organize the discussion in the following order. We start by analysing a particular matter-coupled gauged $\mathcal{N}=1$ supergravity model, with two massive chiral superfields but without three-forms, and we search for supersymmetric AdS vacua. Then, we show how such a standard supergravity theory can be recast into a dual one, which contains the composite gauge three-form vector multiplet and in which the lowest component of the tension superfield $T$, that is $T|_{WZ} = -\frac13 M$, is propagating. This alternative formulation will allow us to couple the system to membranes and construct smooth domain wall solutions.

\subsection{Bulk Lagrangian with two massive chiral superfields}  
\label{sec:2chiralbulk}
The model we study in this section constitutes the bulk Lagrangian for the extended setup which will contain also the membrane. In other words, the supersymmetric AdS vacua we find here can be interpreted as solutions of standard supergravity on the different sides of the membrane added in the following parts.

We start by considering the superspace Lagrangian
\be
\label{2chirals}
\begin{aligned}
{\cal L} =& \alpha \int {\rm d}^4 \theta E \, Z^{r} e^{(6r -2) U} \overline{Z}^{r} 
+ \left( i \int {\rm d}^2 \Theta \, 2 {\cal E} \, X Z  + c.c. \right) 
+ i  \int d^4 \theta E (\overline X - X) e^{-2 U} 
\\
& 
- 3 m \int {\rm d}^4 \theta \, E \,  e^{-2U} 
+ \left( \frac{1}{4g^2} \int {\rm d}^2 \Theta \, 2 {\cal E} \, {\cal W}^2(U) + c.c.  \right)   \, , 
\end{aligned} 
\ee 
where $Z$ and $X$ are chiral superfields, while $U$ is a standard vector superfield which, in the present section, does not contain any gauge three-form. The parameters $\alpha$, $r$ and $m$ are real. Using the standard language of gauged supergravity \cite{Wess:1992cp}, we can identify 
\be
K+\Gamma = 6U  -3 \log \left(m  -\frac13 \alpha Z^{r} e^{6r U} \overline{Z}^{r} - \frac{i}{3} (\overline X - X)   \right) 
\ , \quad W = i X Z \, , 
\ee
where $W$ is the superpotential of the theory. From these expressions, we can calculate the K\"ahler potential and the Killing potential
\be
\label{Potentials}
K=-3 \log \left(m  -\frac13 \alpha Z^{r} \overline{Z}^{r} - \frac{i}{3} (\overline X - X)   \right), \qquad \mathbb{D} = 6+ \frac{18\alpha rZ^r \overline Z^r}{3m - i(\overline X- X)-\alpha Z^r \overline Z^r}.
\ee
It is convenient to express the complex scalars as
\be
\label{comptoreal}
X = \chi + i \phi \ , \quad Z = \rho \, e^{i \theta} \, ,
\ee
where the fields $\rho>0$ and $\theta$ should not be confused with those used in the previous example. In the basis $\Phi^I = \{X,Z\}$, the K\"ahler metric is given by
\be
g_{I \overline J} = \frac{3}{(3m-2\phi-\rho^{2r})^{2}}\left(
\begin{array}{cc}
1 & i \alpha r \rho^{2r} e^{-i\theta}\\
-i \alpha r \rho^{2r} e^{i\theta} & \alpha r^2 (3m -2\phi)\rho^{2(r-1)}
\end{array}\right).
\ee

Supersymmetric AdS vacua can be found by solving the F-term and D-term equations
\be
D_i W=0, \qquad \mathbb{D}=0.
\ee
The solution of these equation is given by the field configuration
\be
\chi=0, \qquad \phi = \frac{3mr}{1-r}, \qquad \rho^{2r} = \frac{3m}{\alpha(1-r)} , 
\ee
and the vacuum energy is 
\be
\langle \mathcal{V} \rangle = \left(\frac{3m}{\alpha(1-r)}\right)^{\frac 1r}\left(\frac{1-r}{rm}\right).
\ee
At this point, the sign of the vacuum energy depends on the choice of $\alpha$, $r$ and $n$. In order to obtain AdS vacua, these parameters can be constrained as follows. First, on the vacuum we ask for consistency that
\be
\det g_{I \overline J}|_{min}=-\frac{(r-1)^2}{27 m^2 r \rho^2 }>0,
\ee
which leads to the condition $r<0$. This, in turn, tells us that we have to require $\frac{m}{\alpha}>0$, in order for $\rho^{2r}$ to be positive. Taking into account these relations, the only way to have a negative vacuum energy is when
\be
\label{paramAdS}
\alpha >0, \qquad r<0, \qquad m>0.
\ee
This region of the parameter space gives supersymmetric AdS vacua.

\subsection{Membrane coupling and BPS flow equations}

We have seen that the model \eqref{2chirals} admits supersymmetric AdS solutions. Now, we would like to introduce a supersymmetric membrane into the setup and study domain walls interpolating between two of such vacua, separated by the membrane. To this purpose, we would have first to trade the standard vector superfield $U$ appearing in \eqref{2chirals} for the composite superfield \eqref{VeryNice} and then we could couple the membrane to the three-form. However, we will follow an alternative path, in order to compare and contrast this example with the previous one. We will start from a model similar to \eqref{compex1} and we will show how to recover \eqref{2chirals} with the three-form correctly coupled to the membrane.

As we explained before, in the previous example the membrane was coupled to the supergravity auxiliary field $M = -3 T|_{WZ}$. Despite the fact that this field was not propagating, it induced non-trivial effects on the equations of motion, which lead to a singularity in the Ricci scalar. In order to avoid this issue and deal with more standard and smooth domain wall solutions, we decide now to let the auxiliary field $M$ propagate. This can be obtained by simply adding a kinetic term for it. Therefore, we start by considering a model similar to \eqref{compex1}, with $S=0$ for simplicity, and with a kinetic term for the chiral superfield $T$. The action is then of the form
\be
\label{Lhd}
\begin{aligned}
S = &  \int d^4 x \left( \alpha \int {\rm d}^4 \theta E \, T^{r} e^{(6r -2) U} \overline{T}^{r} 
+ \left( \frac{1}{4g^2} \int {\rm d}^2 \Theta \, 2 {\cal E} \, {\cal W}^2(U) + c.c.  \right)  
+ {\cal L}_{bd} \right)
\\
& -2 Q \int_{\cal C} {\rm d}^3 \xi \sqrt{-\det h} \left| T \right| 
- Q \int_{{\cal C}}{\cal A}_3 \, ,
\end{aligned}
\ee
where $U$ is now the composite three-form vector multiplet \eqref{VeryNice}. Since the auxiliary field $M$ propagates, this action contains higher derivative interactions, ghost-free nonetheless. 
Indeed, the propagation of the auxiliary field $M$ of old-minimal supergravity is linked to higher curvature terms \cite{Ferrara:1978rk}. 

For convenience, we can dualize \eqref{Lhd} to a standard two-derivatives supergravity action in a manifestly supersymmetric way. 
The result will be precisely the model \eqref{2chirals}, with the composite vector superfiled \eqref{VeryNice}, 
and the system will also be coupled to the supersymmetric membrane. 
To perform the dualization, we consider the parent Lagrangian
\be
\label{Ldualhd}
\begin{aligned}
S = &  \int {\rm d}^4 x \Bigg{(} \alpha \int {\rm d}^4 \theta E \, Z^{r} e^{(6r -2) U} \overline{Z}^{r} 
+ \left(i \int {\rm d}^2 \Theta \, 2 {\cal E} \, X \, (Z - T) + c.c. \right)  
\\
& \qquad \quad  + \left( \frac{1}{4g^2} \int {\rm d}^2 \Theta \, 2 {\cal E} \, {\cal W}^2(U) + c.c.  \right)  
+ {\cal L}_{bd} 
\Bigg{)} 
\\
& -2 Q \int_{\cal C} {\rm d}^3 \xi \sqrt{-\det h} \left| T \right| - Q \int_{{\cal C}}{\cal A}_3 \, ,
\end{aligned}
\ee
where $Z$ is a chiral superfield with the same Weyl weight as $T$, while $X$ is a Lagrange multiplier chiral superfield with vanishing weight. Notice that the chiral superfield entering the membrane tension is still $T$. One can easily check that, by varying $X$, we get $Z=T$ and we recover the higher derivative Lagrangian we started from. To link \eqref{Ldualhd} with the model studied in section \ref{sec:2chiralbulk}, we notice that, using \eqref{TTT}, \eqref{PPP} and up to boundary terms,
\be
\nn
&&  \left(-i \int {\rm d}^2 \Theta \, 2 {\cal E} \, X  T + c.c. \right) 
= \int {\rm d}^4 \theta E (X + \overline X) {\cal P} 
\\
&& = i  \int {\rm d}^4 \theta E (X + \overline X) (\Sigma - \overline \Sigma ) 
= i  \int {\rm d}^4 \theta E (\overline X - X) (\Sigma + \overline \Sigma ) 
\\
\nn
&& 
= i  \int {\rm d}^4 \theta E (\overline X - X) e^{-2 U} \, . 
\ee
Inserting this into \eqref{Ldualhd}, we find eventually 
\be
\begin{aligned}
\label{Ldual2chiral}
S  =&  \int {\rm d}^4 x \Bigg{(} \alpha \int {\rm d}^4 \theta E \, Z^{r} e^{(6r -2) U} \overline{Z}^{r} 
+ \left( i \int {\rm d}^2 \Theta \, 2 {\cal E} \, X Z  + c.c. \right) 
\\
& \qquad \quad + i  \int {\rm d}^4 \theta E (\overline X - X) e^{-2 U} 
+ \left( \frac{1}{4g^2} \int {\rm d}^2 \Theta \, 2 {\cal E} \, {\cal W}^2(U) + c.c.  \right)  
+ {\cal L}_{bd} \Bigg{)} 
\\ 
& -2 Q \int_{\cal C} {\rm d}^3 \xi \sqrt{-\det h} \left| T \right| - Q \int_{{\cal C}}{\cal A}_3 \, . 
\end{aligned} 
\ee
This is precisely the gauged supergravity model \eqref{2chirals} with two chiral superfields, but now the vector superfield $U$ is given by \eqref{VeryNice} and the three-form is consistently coupled to a membrane. In particular, this action admits the same supersymmetry AdS vacua studied in section \ref{sec:2chiralbulk}.  We stress once more that, similarly to the previous example, the chiral superfield entering the Nambu--Goto term of the membrane tension has as lowest component the auxiliary field $M$.

The component form of the Lagrangian \eqref{Ldual2chiral}, under the assumption \eqref{novect} and in the field basis \eqref{comptoreal} with $\chi=0$, is given by
\begin{equation}
\label{LagSmooth}
    \begin{aligned}
    e^{-1}\mathcal{L} &= \frac16 (\alpha \rho^{2r}+2\phi)R - \alpha r^2 \rho^{2(r-1)}\partial_m (\rho e^{i\theta}) \partial^m (\rho e^{-i\theta}) \\
    &+\frac19 (\alpha \rho^{2r}+2\phi)M\overline M - \frac13 \alpha r \rho^{2r-1}e^{-i\theta} F^Z M -\frac13 \alpha r \rho^{2r-1}e^{i\theta} \overline F^Z \overline M \\
    &+\frac i3 M F^X - \frac i3 \overline M \overline F^X + \alpha r^2 \rho^{2(r-1)}F^Z \overline F^Z+\phi \rho e^{i\theta}\overline M + \phi \rho e^{-i \theta} M \\
    &- 2\phi {\rm Re}F^Z +i \rho e^{i\theta}F^X - i \rho e^{-i\theta}\overline F^X\\
    &+\frac12 (\alpha (6r-2)\rho^{2r}-4\phi) {\rm D} + \frac{1}{2g^2}{\rm D}^2+e^{-1}\mathcal{L}_{bd}\\
    &+e^{-1}\int {\rm d}^3\xi\, \delta^{(4)}(x-X)\left(-\frac23 Q |M| \sqrt{-\det h} - \frac13 Q \tilde \epsilon^{ijk}\partial_i X^m \partial_j X^n \partial_k X^p C_{mnp}\right),
    \end{aligned}
\end{equation}
where D is the composite expression \eqref{Dcomp}, with again $v_m=b_m=0$. The equations of motion of the auxiliary fields, using the condition \eqref{D=0}, result in
\begin{align}
\label{EOMM}
\delta F^X: \,\,\,\, \qquad M =& - 3 \rho e^{i\theta},\\
\delta M:\quad -i F^X =&\, \left(\alpha\rho^{2r+1}-\frac{\phi\rho}{r}\right)(1-3r)e^{-i\theta}\\
&+ e^{-1}Q\int d^3\xi \delta^{(4)}(x-X)e^{-i\theta}\sqrt{-\det h},\\
\label{EOMFZ2}
\delta F^Z: \,\,\,\qquad F^Z=& -\frac{\rho^2}{r}+ \frac{\phi}{\alpha r^2 \rho^{2(r-1)}},
\end{align}
while the variation of the real scalar $\phi$ gives
\be
\label{EOMC32}
\delta \phi: \qquad *{\rm d}C_3 = 
-\rho^2 \left(3-\frac 1r\right)- \frac{\phi}{\alpha r^2 \rho^{2(r-1)}}.
\ee
It is important to notice that now the auxiliary field $M$ does not contain any Dirac delta term anymore. This will allow us to find a continuous profile for the warp factor.

So far the discussion has been quite general, but in order to proceed we prefer to make some additional assumptions, to simply the setup. First, we recall that we are working with the metric ansatz \eqref{metric4d} and that all fields are assumed to depend at most on $z$, in particular
\be
\rho = \rho(z), \qquad \phi = \phi(z).
\ee
We further assume that 
\be
e^{i\theta}\equiv 1 , \qquad  M= \overline M \equiv \mathcal{M}, \qquad F^Z = \overline F^Z \equiv \mathcal{F}, \qquad -iF^X =i\overline F^X \equiv \mathcal{H} \, , 
\ee
which, together with \eqref{EOMM}, imply 
\be
{\rm sgn}(\mathcal{M}) = - 1  \, . 
\ee
Finally, we choose the parameters
\be
\alpha = 1,\qquad r=-1,
\ee
in accordance with \eqref{paramAdS}, but we do not fix $m$. From now on, we will always work in the static gauge \eqref{staticgauge}.

As we explained in the previous example, in order for (part of) the bulk supersymmetry to be preserved by the membrane, we have to require that the parameter $\xi_\alpha$, at $z=0$, is identified with the kappa-symmetry parameter $\kappa_\alpha$. In particular, the latter satisifes \eqref{kappa} and therefore we impose
\be
\xi_\alpha 
= \frac{T}{|T|} \left( \frac{i \tilde \epsilon^{ijk}}{3! \sqrt{-\det h}} \epsilon_{abcd} E^b_i E^c_j E^d_k \sigma^a_{\alpha \dot \alpha} \right) \Big{|}_{z=0} \overline \xi^{\dot \alpha} =- \frac{i M}{|M|}\Big{|}_{z=0} \sigma^3_{\alpha \dot \alpha} \overline \xi^{\dot \alpha} = ie^{i\theta}\Big{|}_{z=0} \sigma^3_{ \alpha \dot \alpha} \overline \xi^{\dot \alpha}. 
\ee
By comparing with \eqref{xiproj}, we obtain then 
\be 
e^{i\beta}= i e^{i\theta}|_{z=0} \equiv i \, . 
\ee

After all of these assumptions and simplifications, the BPS equations \eqref{BPS1} and \eqref{BPS2} become
\begin{align}
\label{BPSrho2}
\dot \rho &=- \mathcal{F},\\
\label{BPSphi}
\dot \phi &= -\mathcal{H},\\
\label{BPSA2}
\dot A &= \frac{\mathcal{M}}{3},
\end{align}
where the variation of the auxiliary fields and of the scalar $\phi$ are given respectively by
\begin{align}
\label{Mrho2}
\mathcal{M} &= - 3 \rho,\\
\mathcal{H} &= Q \, \delta(z) + 4\phi\rho + 4 \rho^{-1} ,\\
\label{EOMF2}
\mathcal{F} &= \rho^2 +\phi \rho^4,\\
\label{eomphi2}
*{\rm d}C_3 &= -4\rho^2-\phi\rho^{4}.
\end{align}
The equation of motion \eqref{eomXm} reduces to
\be
\label{eomXm2}
*{\rm d} C_3 =  e^{-3A}\partial_z \left(\rho \sqrt{-\det h}\right) =  e^{-3A}\partial_z \left(\rho e^{3A}\right) =  \dot \rho  + 3 \rho \dot A,  
\ee
which is in agreement with \eqref{BPSA2} and \eqref{eomphi2}, while the equation of motion of the three-form, under the assumption \eqref{D=0}, leads to
 \be
 \label{constrphi}
 3 m  = 2 \phi + 4 \rho^{-2} \ , \quad  m = c_o -  \frac23 Q \Theta(z) \, . 
\ee

As a check, we can verify that the condition \eqref{D=0} is satisfied along the flow. Indeed, for $v_m=b_m=0$ and using \eqref{Ricci}, together with \eqref{Mrho2} and \eqref{eomXm2}, we have
\be
{\rm D}\equiv \frac16 R + *{\rm d}C_3 + \frac19 \mathcal{M}^2 = \left(\partial_z +2\dot A -\frac{\mathcal{M}}{3}\right)\left(\dot A - \frac{\mathcal{M}}{3}\right) = 0,
\ee
which vanishes along \eqref{BPSA2}. By using \eqref{Tcomp}, \eqref{EOMM} and \eqref{EOMC32}, one can check also that
\be
F^T = - *{\rm d}C_3 - \frac13 M \overline M = F^Z,
\ee
where $F^Z$ is given in \eqref{EOMFZ2}. This is expected from the very form of the superspace Lagrangian \eqref{Ldualhd}, from which one can derive that $T=Z$, on-shell.

Before solving the BPS equations, we would like to verify that the Ricci scalar has 
the standard behavior we expect from a domain wall solution. 
The Lagrangian \eqref{LagSmooth} is not in Einstein frame, since its effective Hilbert--Einstein term is
\be
{\cal L}_{\text{HE, on-shell}} = - \frac12 \rho^{-2} e R \, . 
\ee
Notice here that the effective $M_P^2$ we discussed in \eqref{MP2} can be identified as $n=\rho^{-2}$. 
The appropriate Weyl rescaling is this time $g_{mn} = \rho^2 g_{mn}^E$ and the Einstein frame metric is explicitly 
\be
\label{metric4dE2}
ds^2_{E} = \rho(z)^{-2}\left[ e^{2A(z)}(-dt^2+dx^2+dy^2) + dz^2\right] \, . 
\ee 
We notice that, in this example, the Weyl rescaling is performed with a continuous function, which however does not have continuous derivatives.
The Ricci scalar associated to \eqref{metric4dE2} is then
\be
\begin{aligned}
R_E &= \rho^2 R+6\rho^2\left(2 \left(\frac{\dot\rho}{\rho}\right)^2 - \frac{1}{\rho}(3\dot A\dot \rho + \ddot \rho)\right)\\
&=6\left(2\rho^4 + 2\rho^2 \dot \rho + 2 \dot \rho^2 - \rho \ddot \rho\right),
\end{aligned}
\ee
where $R$ is given in \eqref{Ricci}. 
The most singular behavior on the membrane source is given by the last term and it is of the type
\be
R_E|_{z=0} \sim \delta(0)+\dots 
\ee
This is the standard behavior expected for the curvature on the membrane, in such kind of models.

In addition, it is easy to verify that, along the flow, the equations of motion for the scalar $\rho$ and the traced Einstein equations for the metric that can be derived from \eqref{LagSmooth} are also satisfied in this example, even on the membrane. This was not the case for the previous model and provides a non-trivial cross-check for the self-consistency of this solution.

\subsection{Domain wall solutions}

Finally, we present the solution of the BPS flow equations \eqref{BPSrho2}, \eqref{BPSphi} and \eqref{BPSA2}, namely
\begin{align}
\label{flowflow}
    \dot\rho &= -\rho^2 - \phi \rho^4,\\
    \dot\phi &= -Q \, \delta(z) - 4\phi\rho - 4 \rho^{-1},\\
    \dot A &= -\rho.
\end{align}
To simplify the problem, one can notice that the second equation in \eqref{flowflow} is identically satisfied by \eqref{constrphi}. Therefore, we can just restrict our attention to
\begin{align}
\label{BPSrho22}
    \dot \rho &= \rho^2 - \frac32 \rho^4\left(c_0 -\frac23 Q \Theta(z)\right),\\
    \label{BPSA22}
    \dot A &= -\rho.
\end{align} 
One can find analytic solutions to these equations and verify that away from the membrane we find the supersymmetric AdS vacua we discussed in section \ref{sec:2chiralbulk}, with asymptotic values for $\rho$ and $R_E$ given by 
\be
\rho^{-2}|_{\rm asymptotic} = 3m / 2 \ , \quad R_E|_{\rm asymptotic} = \frac{16}{3 m^2} \, .  
\ee

\begin{figure}[ht]
\centering 
  \includegraphics[scale=.85]{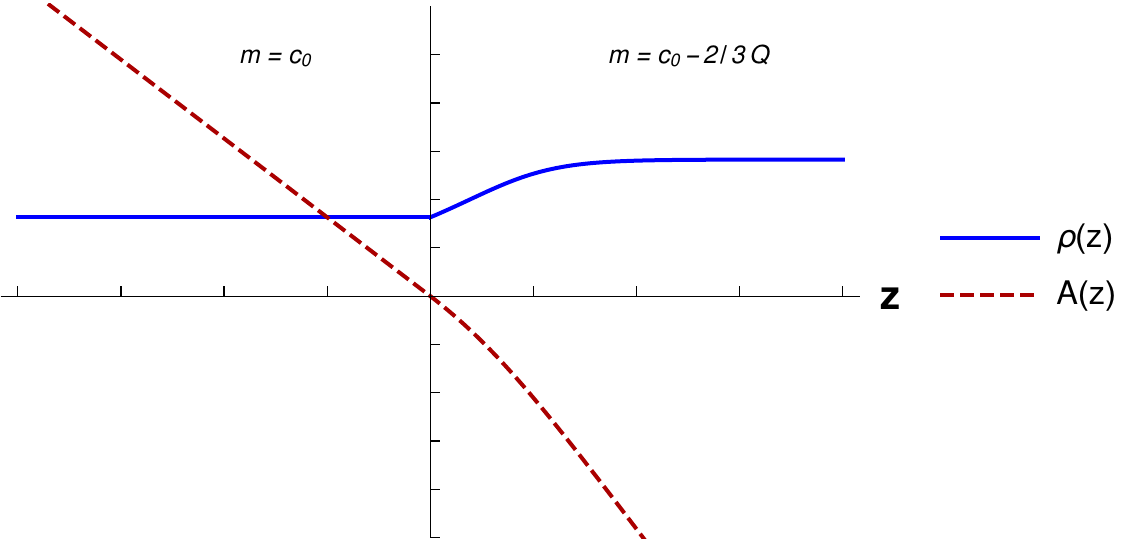}
\caption{ 
{\it Plot of a numerical solution of \eqref{BPSrho22} and \eqref{BPSA22} for the scalar $\rho(z)$ and the warp factor $A(z)$, with parameters $c_0  =1$ and $Q=1$. We see that both fields are continuos everywhere along the $z$-direction. On the right hand side we have asymptotically supersymmetric AdS.} \label{plotz2}} 
\end{figure}

It is however more convenient for our presentation to continue with numerical solutions. We will use the values 
\be
c_0 = 1 \ , \quad Q = 1 \, . 
\ee 
We found a numerical solution to the system of the differential flow equations, which is reported in figure \ref{plotz2}. 

One can see that, contrary to the previous example, both the profiles of the scalar fields are continuous. In addition we see that the scalar $\rho$ has a non-trivial flow only on the one side of the membrane and it also goes asymptotically to its value characterized by the appropriate supersymmetric AdS vacuum on that side. 
In particular, for our solution the asymptotic values of $\rho$ are 
\be
\rho|_{\rm asymptotic} = \left\{\begin{array}{cc}
     \sqrt{2/3} &{\rm for}  \ z<0 ,\\[.3cm] 
     \sqrt{2}   &{\rm for} \ z \, \to \, +\infty .\\
\end{array}\right.  
\ee 

\begin{figure}[ht]
\centering 
  \includegraphics[scale=.7]{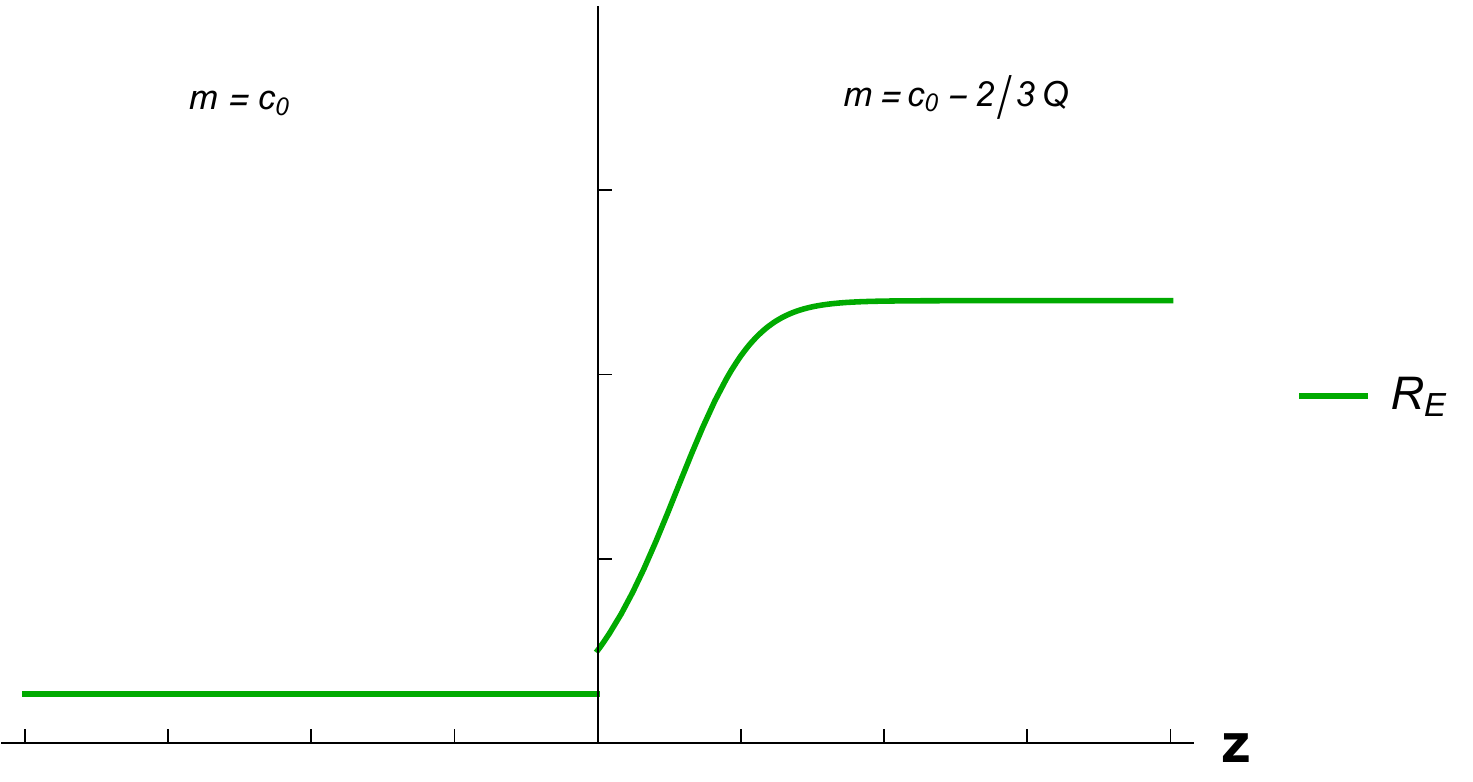}
\caption{ 
{\it Plot of the behavior of the Einstein frame Ricci scalar for our numerical solution. Note that it has a monotonic behavior.} \label{plotz33}} 
\end{figure}

The metric background also goes asymptotically to AdS with values for the Einstein frame curvature 
\be
R_E|_{\rm asymptotic} = \left\{\begin{array}{cc}
     16/3 &{\rm for}  \ z<0 ,\\[.3cm] 
     48   &{\rm for} \ z \, \to \, +\infty .\\
\end{array}\right.  
\ee
For completeness, we present also the profile of the Einstein frame Ricci scalar $R_E$ along $z$, which gives information on the cosmological constant in figure \ref{plotz33}. 
For our example, with $c_0=1$ and $Q=1$, the exact form of the curvature is 
\be
R_E|_{c_0 = Q=1} = -3\rho^4(-8 + 9 m^2 \rho^4 + 2 \rho \,\delta(z)), 
\ee
where we see the Dirac function type of singularity on the membrane source. 

\section{Discussion}

In this work, we presented a new vector multiplet in supergravity, which contains a gauge three-form inside its D-term component field. We proposed also a superspace Lagrangian, describing its interactions. This Lagrangian is super-Weyl invariant off-shell, but the latter invariance is spontaneously broken when the gauge three-form is integrated out and the Planck scale is introduced dynamically into the theory. As we showed explicitly, this theory can be recast into the more familiar Lagrangian for the standard Fayet--Iliopoulos model of D-term breaking in supergravity. At this stage, we commented also on the fact that pure Fayet--Iliopoulos terms in supergravity are in tension with the Weak Gravity Conjecture. Finally, we studied the couplings to charged chiral superfields and to effective membranes in four dimensions, and we constructed domain walls in two different examples, one of those containing also higher derivative interactions. In these models, the vacua from the left and the right of the membrane source are characterized by a different Planck mass. 

A possible interpretation of the irregularities encountered in the first domain wall profile is that the associated system does not admit a supersymmetric ground state at all. However, we refrained to make such a statement in the main part of the work, since we preferred to remain on the safe side, without drawing any conclusion. It is nevertheless interesting to comment further on this hypothesis. Assuming this to be true, the model of section \ref{sec:smoothdw} might then be among the simplest generalizations of the previous irregular solution admitting a consistent BPS ground state. This may imply that whenever the membrane tension depends on an auxiliary field, the existence of the corresponding BPS ground state might be jeopardized, unless higher derivative corrections are introduced. In this respect, knowing the off-shell completion of a given theory would be important in order to understand how to couple consistently to extended objects.

An aspect of our construction that we find quite surprising is the fact that we start from an off-shell superspace description of a locally supersymmetric Maxwell theory and eventually we recover gravity dynamically, when going on-shell. In this sense, we are indeed giving a new off-shell completion of a supersymmetric theory with just an abelian vector multiplet, which is on-shell equivalent to the standard supersymmetric Einstein--Maxwell model. This may be also relevant for the search of off-shell completions in supergravity theories with more supercharges.

Finally, an interesting development would be to employ this new three-form multiplet as part of low energy effective theories coming from flux compactifications of string theory, extending earlier works \cite{Ceresole:2006iq,Carta:2016ynn,Farakos:2017jme,Bandos:2018gjp,Herraez:2018vae,Lanza:2019xxg}, and in particular to study the compatibility with \cite{Lanza:2019xxg} 
and the relation to \cite{Hartong:2009az}. 
In string compactifications the four-dimensional Planck mass is related to the volume of the internal manifold. In situations in which the moduli can be stabilized, see for example \cite{DeWolfe:2005uu}, the volume would be given then by flux numbers. In this respect, our construction can be seen as a toy model for these more fundamental scenarios, since the Planck mass is given by the flux $n$.

\section*{Acknowledgements} 

We thank David Andriot, Alex Kehagias, Stefano Lanza, Luca Martucci, Lorenzo Papini, Dmitri Sorokin, Magnus Tournoy, Thomas Van Riet and Timm Wrase for discussions. 
The work of NC is supported by an FWF grant with the number P 30265. 
The work of FF is supported by the KU Leuven C1 grant ZKD1118C16/16/005. 
GT thanks the ITF of KU Leuven for the kind hospitality while this article was prepared.




\providecommand{\href}[2]{#2}\begingroup\raggedright\endgroup

\end{document}